\documentclass[11pt,english,fleqn]{article}
\usepackage[T1]{fontenc}
\usepackage{authblk}
\usepackage[utf8]{inputenc}
\usepackage[english]{babel}
\usepackage{graphicx}
\usepackage[round, sectionbib,authoryear]{natbib}     
\usepackage[labelformat=simple]{subcaption}
\usepackage{tabularx}
\usepackage{multirow}
\usepackage{graphicx}
\usepackage{psfrag}
\usepackage{amsmath}
\usepackage{amssymb}
\usepackage{nicefrac}
\usepackage[]{algorithm2e}
\usepackage{fullpage}
\usepackage{listings}
\usepackage{enumerate}
\usepackage{wrapfig}
\usepackage{float}
\usepackage{parskip}
\usepackage{lipsum}
\usepackage{listings,color}
\usepackage[font=small,labelfont=normal,labelsep=default]{caption}
\usepackage[nodisplayskipstretch]{setspace}
\usepackage{epstopdf}
\AppendGraphicsExtensions{.tif}
\usepackage[colorlinks=true,urlcolor=blue,citecolor=red,linkcolor=red,bookmarks=true]{hyperref} 
\usepackage{cleveref}
\usepackage{pstricks,pstricks-add,pst-3d,pstricks-add,pst-grad,multido}

\definecolor{halfgray}{gray}{0.55} 
\definecolor{webgreen}{rgb}{0,.5,0}
\definecolor{webbrown}{rgb}{.6,0,0}
\definecolor{BlueLUH}{cmyk}{1.0,0.7,0,0}
\colorlet{LightBlue}{BlueLUH!20!white}
\colorlet{DarkBlue}{BlueLUH!80!black!20}

\title{Dynamics of two air bubbles rising in a shear-thinning fluid}

\author[1]{Purushotam Kumar\thanks{pkumar8@illinois.edu}}
\affil[1]{Corning Incorporated, Manufacturing Technology \& Engineering, Corning, NY 14830}
\author[2]{Surya P. Vanka\thanks{spvanka@illinois.edu}}
\affil[2]{Mechanical Science \& Engineering, University of Illinois Urbana-Champaign, Urbana, IL 61801}

\date{\today}
\providecommand{\keywords}[1]{Keywords: #1}

\begin{document}

\maketitle
\onehalfspacing

\begin{abstract}	
	{\it In this paper, we have studied the three-dimensional dynamics of two equally sized air bubbles rising in a shear-thinning fluid. We have used the combined level set and volume of fluid (CLSVOF) method to track interface, maintain mass balance and estimate the interface curvature. Additionally, we have incorporated a Sharp Surface Force Method (SSF) for surface tension forces. This method significantly suppressed the spurious velocities commonly observed with the conventional volume of fluid method and the Continuum Surface Force (CSF). The algorithm is implemented in an in-house code called CUFLOW and runs on multiple GPUs platform.
		
	We have explored the effects of fluid rheology on the three-dimensional dynamics of two in-line bubbles. Two power-law indices (0.5 and 1) are investigated to highlight differences in shear-thinning and Newtonian fluids. For a range of parameters examined here, bubbles motion in a shear-thinning fluid is seen to be unsteady with significant shape oscillations. Further, we have examined the rise velocity, droplets rise path, transient shapes and found that modification of viscosity by the motion of the leading bubble changes the dynamics of the trailing bubble.
}  
\end{abstract}


\section{Introduction}
\label{sec:introduction}
A gas bubble rising in a liquid medium is one of the simplest two-phase flow problems and addresses several important interactions between two phases. This problem has been previously studied to understand lift \cite{Morsi1972particle,Kurose2001drag_lift} and drag \cite{Magnaudet1998drag} forces acting on a bubble, lateral migration of the bubble in a co-flow \cite{ADOUA2009Lift_force_reversal}, and bubble-wall and bubble-wake interactions. A large number of researchers, including \citet{Allen1900,Hadamard1911,Rybczynski1911,Bond1927,Moore1959} have investigated the bubble deformation, rise path, terminal velocity, and confinement effects through experiments, theoretical modeling, and numerical simulations.

Dynamics of a rising bubble are a function of the E\"otv\"os number ($Eo = \Delta \rho g d^2 \sigma^{-1}$, also referred as Bond number $Bo = \rho_l g d^2 \sigma^{-1}$), Morton number $(Mo = g \mu_o^4 \rho_l^{-1} \sigma^{-3})$, confinement ratio $(C_r = Wd^{-1})$, and other external forces such as a magnetic field \cite{Jin2016} and variable viscosity \cite{Kumar2015non_newtonian}. An exhaustive collection of experimental data for a bubble rising in an unconfined medium has been summarized in works of \citet{Grace1973,Grace1976,Bhaga1981}, available as nomogram charts of the terminal Reynolds and bubble shape for combinations of Bond and Morton numbers. In addition, the proximity of the bubble to a wall alters the bubble rise velocity and shape and increases the shear force at the gas-liquid interface \cite{Figueroa-Espinoza2008,Kumar2015confinement}. Detailed reviews of bubble dynamics in Newtonian fluids are available in the works of \citet{Clift1978,DeKee2002}.

The dynamics of bubbles in Newtonian fluids have been widely studied. However, many fluids have non-Newtonian behavior and are used in the chemical and pharmaceutical industries. For instance, the fermentation of liquor, sugarcane processing, ethanol extraction, aeration of the membrane is typical shear-thinning applications. Bubble rise in non-Newtonian fluids is more complex because the bubble-generated velocities subsequently change the local fluid viscosity due to the shear. Further, the nature of the fluid (Bingham, thixotropic, shear-thinning, shear-thickening, etc.) is important in modifying the fluid viscosity in the bubble vicinity and thereby the local shear stress values. Thus, a large parameter space is available, even for a single bubble in a non-Newtonian fluid, which provides rich and complex flow physics regions. Previously, several efforts have been made towards understanding such bubble motion in viscoelastic and inelastic non-Newtonian fluids. Some example works are by \citet{Astarita1965,Leal1971,Carreau1974,Zana1978,Kawase1981,Dekee1986,Gummalam1988,Dekee1990,Manjunath1992,Miyahara1993,Liu1995,Rodrigue1996,Li1997,Chhabra1998,Gauri1999,Dubash2007Propagation,Tsamopoulos2008Steady,Sikorski2009Motion,Kishore2013,Tripathi2015} for bubble motion in shear-thinning, shear-thickening and Bingham fluids. The papers of \citet{Kulkarni2005,Chhabra2006} provide good summaries of the important results. \par

The dynamics of multiple bubbles rising has been studied by several authors \citep{Abbassi2018,Gumulya2017,Watanabe2006,yuan_prosperetti_1994,legendre_magnaudet_mougin_2003,Yu2011LBM,KATZ1996239,RUZICKA20001141} and have reported that flow structure around the bubble has an important impact on the interaction between leading and trailing bubbles. They have reported that wake in front of a trailing bubble plays an important role in reducing the drag around it. Due to reduced drag, the trailing bubble accelerates until it coalesces with the leading bubble. Along with these, there are other studies; however, the majority of the studies for multiple bubble rise are conducted in the Newtonian fluids.  \par 
In the present paper, we have briefly discussed a numerical technique to simulate two-phase flows and used it to study the three-dimensional deformation and rise of two equally sized bubbles in shear-thinning and Newtonian fluids for a given bubble size and center to center distance. Despite a reasonably large number of previous studies, there are very few studies that have examined the dynamics of in-line bubbles rising in non-Newtonian fluids. Various quantities such as the bubble shape, rise velocity, and rise paths are investigated. The algorithm used for the computations is presented in the numerical method section. The problem description is provided in the computational details section \ref{sec:computational_details}. In the results and discussion section \ref{sec:results}, we present the results of the simulations and discuss the critical findings. A summary of the present findings is given at the end. \par

\section{Numerical Method}
\label{sec:numerical_method}

\hspace{3mm} We have developed a numerical procedure that solves the relevant governing equations on a fixed Eulerian collocated grid. The procedure consists of algorithms for interface capturing, the inclusion of surface tension forces, and the solution of Navier-Stokes equations. The volume of fluid (VOF) method of Hirt and Nichols \cite{Hirt1981} is used for interface capturing. The truncated interface between the gas and liquid in a control volume is represented as an oblique plane, and the interface normal, and curvature are computed using a smoothed liquid volume fraction and the height function method (Rudman \cite{Rudman1998} and Cummins et al. \cite{Cummins2005}), respectively. We have incorporated the surface tension force in Navier-Stokes equations using a sharp surface force (SSF) and pressure balance methods (Francois et al.\cite{Francois2006}, Wang and Tong \cite{Wang2010}). The Navier-Stokes equations are solved using the fractional step method. The numerical procedure has been previously used to study the effects of duct confinement on dynamics of bubble rising in a square duct \cite{Kumar2015confinement,Kumar2015numerical} and the effects of magnetic field on bubble dynamics \cite{Jin2016mhd_bubble}.  \par 
\subsection*{Governing Equations}
\hspace{3mm} We assume that both the gas and liquid are isothermal and incompressible. A single fluid approach with a method to capture the interface between gas and liquid is used. The combined governing equations for both fluids are given by the following equations:
\begin{alignat}{1}
	\label{eqn:continuity}
	\nabla \cdot  \mathbf{u} = 0
\end{alignat}
\begin{equation}
	\label{eqn:momentum_equation}
	\begin{aligned}
		\frac{\partial \left(\rho \mathbf{u} \right)}{\partial t} + \nabla \cdot \left( \rho \mathbf{uu} \right) =  -\nabla p & + \nabla \cdot \left( \mu \left[ \nabla \mathbf{u} + \nabla \mathbf{u}^{T} \right] \right) \\ & + \rho \textbf{g} + \sigma \kappa \mathbf{n} \delta \left( \mathbf{x} - \mathbf{x}_f \right)
	\end{aligned}
\end{equation}
In the above equations, $\mathbf{u}$ is the fluid velocity, $p$ is pressure, $\rho$ is density, $\mu$ is dynamic viscosity, $\sigma$ is surface tension coefficient, $\kappa$ is interface curvature, $\mathbf{n}$ is interface normal, $\delta$ is the delta function, $\mathbf{x}$ is the spatial location where the equation is solved, $\mathbf{x}_f$ is the position of the interface and $\mathbf{g}$ is the acceleration due to gravity. \par
\hspace{3mm} Since the free surface is captured by the volume of fluid method, a time evolution of VOF function (liquid volume fraction, $\alpha$) given by
\begin{alignat}{1}
	\label{eqn:vof_eqn}
	\frac{\partial \alpha}{\partial t} + \textbf{u} \cdot \nabla \alpha = 0 
\end{alignat}
is solved. The geometry construction method (Rider and Kothe \cite{Rider1998}) coupled with a second-order operator split method (Noh and Woodward \cite{Noh1976}, Li \cite{Li1995}, Ashgriz and Poo \cite{Ashgriz1991} and Francois et al. \cite{Francois2006}) is used to solve the evolution equation. The mixture density and viscosity are calculated as a linear weighting of the individual phase values, as
\begin{alignat}{2}
	\label{eqn:density_viscosity}
	\rho &= \alpha \rho_l+ (1 - \alpha)\rho_g \\
	\mu  &= \alpha \mu_l + (1 - \alpha)\mu_g 
\end{alignat}
The subscripts $l$ and $g$ denote gas and liquid phases, respectively. The last term in eq. \eqref{eqn:momentum_equation} is the surface tension force, and $\sigma \kappa$ is the pressure jump at the free surface. In our current algorithm, the jump condition is enforced explicitly by representing the surface tension force as a pressure gradient given by
\begin{alignat}{1}
	\label{eqn:jump_condition}
	\sigma \kappa \mathbf{n} \delta \left( \mathbf{x} - \mathbf{x}_f \right)  = - \nabla \tilde{p} 
\end{alignat}
Where $\tilde{p}$ is the pressure solely due to the surface tension force at the interface. An elliptic equation for $\tilde{p}$ is derived from continuity and momentum equations,
\begin{align}
	\label{eqn:ppe_surface_tension}
	\nabla \cdot \left( \frac{\nabla \tilde{p}}{\rho} \right) = 0
\end{align}
\hspace{3mm} The jump condition at the interface is enforced by using eq. \eqref{eqn:jump_condition} as a source term in the solution of eq. \eqref{eqn:ppe_surface_tension}. This ensures the exact difference in pressure at the interface due to the surface tension. In our work, this elliptic equation for $\tilde{p}$ is efficiently solved using a multigrid accelerated red-black SOR relaxation scheme implemented on a GPU. Using this method, the spurious velocities reduces to machine zero for a static bubble with exact analytical curvature and to very small values when the curvature is numerically computed. Admittedly, there are several methods to include the surface tension force in the Navier-stokes (Renardy and Renardy \cite{Renardy2002}, Sussman et al. \cite{Sussman2003,Sussman2007}, Gueyffier et al. \cite{Gueyffier1999}); however, in our experience, the current method is more accurate for reduction of spurious velocities and relatively easier to implement on GPUs. \par
\subsection*{Solution Procedure}
The following steps outline solution of the governing equations,
\begin{enumerate}
	\item Initialize the solution with initial liquid fraction ($\alpha^n$) and velocities ($\mathbf{u}^n$)
	\item Calculate the density and viscosity at $n^{th}$ time step using,
	\begin{alignat*}{1}
		\rho^n = \alpha^n \rho_l + (1 - \alpha^n)\rho_g \\
		\mu^n  = \alpha^n \mu_l  + (1 - \alpha^n)\mu_g
	\end{alignat*}
	\item Solve the interface tracking equation (using $\mathbf{u}^n$) to obtain $\alpha^{n+1}$
	\item Calculate the density ($\rho^{n+1}$) and viscosity ($\mu^{n+1}$) using $\alpha^{n+1}$
	\item Calculate the interface curvature using the height function method
	\item Solve the pressure $(\tilde{p})$ using sharp surface force method
	\begin{alignat*}{1}
		\nabla \cdot \left( \frac{\nabla \tilde{p}}{\rho}\right)^{n+1} \hspace{5mm} \text{with} ~\nabla \tilde{p} = - \left( \sigma \kappa \mathbf{n} \delta \left( \mathbf{x} - \mathbf{x}_f \right) \right)^{n+1}
	\end{alignat*}
	\item Compute the pressure gradient term $\left( \frac{\nabla \tilde{p}}{\rho}\right)^{n+1}$ at cell faces
	\item Compute the convection term $(\nabla \cdot \rho \textbf{uu})$ using geometry construction method
	\item Compute the intermediate velocities at cell centers as,
	\begin{equation*}
		\begin{aligned}
			\rho_c^{n+1}\hat{\mathbf{u}}_c = & \rho_c^n \mathbf{u}_c^n - \Delta t \sum_f \mathbf{u}_c^n \left(\sum_{k=1}^{2} \rho_k^n \alpha_k^n\right) A_f \\ 
			&+ \Delta t \sum_f \mu_f^n \left(\nabla \mathbf{u}^n + \nabla^T \mathbf{u}^n\right)_f \cdot \mathbf{n}_f + \rho_c^{n+1} \left(\frac{\nabla \tilde{p}}{\rho}\right)_{f \rightarrow c}^{n+1}
		\end{aligned}
	\end{equation*}
	\item Compute the intermediate velocities at cell faces as, 
	\begin{alignat*}{1}
		\hat{\mathbf{u}}_f = \hat{\mathbf{u}}_{c \rightarrow f}
	\end{alignat*}
	\item Compute pressure $(p)$ using,
	\begin{alignat*}{1}
		\nabla \cdot \left( \frac{\nabla p}{\rho} \right)^{n+1} = \frac{\nabla \cdot \hat{\mathbf{u}}_f}{\Delta t}
	\end{alignat*}
	\item Correct the velocity field for divergence free condition using,
	\begin{alignat*}{1}
		\mathbf{u}_c^{n+1} = \hat{\mathbf{u}}_c - \Delta t \left( \frac{\nabla p}{\rho} + \frac{\nabla \tilde{p}}{\rho} \right)_{f \rightarrow c}^{n+1}
	\end{alignat*}
\end{enumerate}
Where, $_{c \rightarrow f}$ represents the linear averaging procedure to compute face value (eg. $\rho_f$, $\mu_f$, $\mathbf{u}_f$) from its cell-centered values. The intricate details of the numerical algorithm are described in Kumar and Vanka \cite{Kumar2015confinement}. 

The entire algorithm has been programmed on the CUDA-Fortran platform supported by the Portland Group (PGI) Fortran compiler to run on multiple graphics processing units (GPU). Previously, we have validated the multiple GPU implementation \citep{Kumar2015numerical,Vanka2016Single,Kumar2016thesis} to study confinement effects on bubble dynamics \citep{Kumar2015confinement}, two-phase flows at T-juctions \citep{horwitz2012simulations,horwitz2013simulations,kumar2013three}, bubble dynamics in non-Newtonain fluids \citep{ Kumar2015non_newtonian,Kumar2019AJKFluids}, droplet dynamics in square duct \citep{Horwitz2014_lbm,Horwitz2019AJKFluids}, Argon bubble rising in liquid steel \citep{Vanka2015APS_DFD,Jin2016mhd_bubble,Kumar2022APS_DFD,kumar2023_MHD_inline_Bubble} and turbulent bubbly flow \citep{Vanka2016APS_DFD,kumar2021_bubbly_flow}. \par

\section{Computational Details}
\label{sec:computational_details}
\subsection*{Problem Setup}

\begin{figure}[h]
	\begin{center}
		\includegraphics[height=0.4\textwidth]{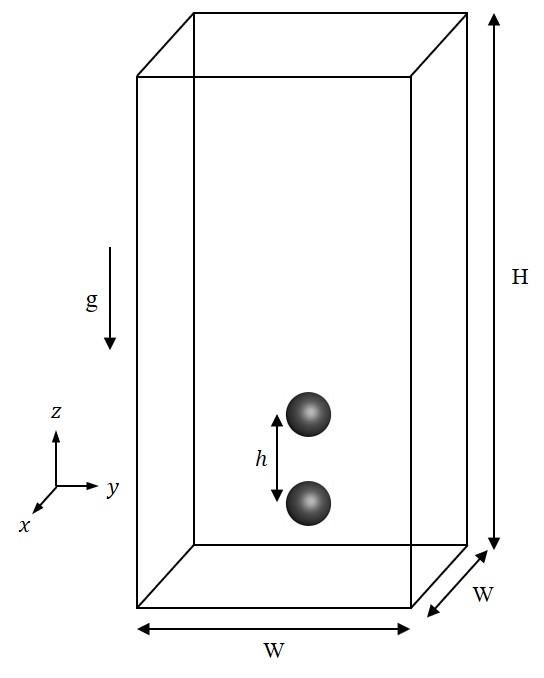}
	\end{center}
	\caption{Initial position of the bubble}
	\label{fig:problem_statement}
\end{figure}

Figure \ref{fig:problem_statement} shows the geometry of the problem considered. The dimensions of domain are $4d \times 4d \times 24d$ along $x$, $y$ and $z$ directions respectively. The height was chosen such that the bubble in a Newtonian fluid can reach a steady velocity at the end of the domain. All boundaries of the domain are considered to be no-slip and no-penetration walls. The trailing bubble is initially placed two diameters above the bottom boundary in the center of the duct cross-section. The leading bubble is placed according to the bubble's center to center distance of the problem (in this case $4d$). The gravitational force acts downwards along the negative z-axis. Based on a systematic grid refinement study, we have selected $32$ control volumes per bubble diameter as an adequate resolution, with a grid of $128 \times 128 \times 768$ ($\approx$ 13 million) control volumes current study.

The liquid and gas densities are taken as 1000 and 1.2 $kg/m^3$. The zero shear viscosity $(\mu_0)$ and time constant $(\lambda)$ of the liquid are $0.014$ $Pa.s$ and 1 $s$ respectively. The power-law index is varied from each case. The dynamic viscosity of the gas $(\mu_g)$ is taken as $1.8 \times 10^{-5} Pa.s$. The surface tension $(\sigma)$ between liquid and gas phase is prescribed a value of $0.072 N/m$\footnote{In practice there is variation in the surface tension due to addition of polymers for preparation of shear-thinning and shear-thickening fluids but the variation is relatively less significant for low concentration of polymers \cite{Ohta2003_shear_thinning,Zhang2010,Ohta2012Shear_thickening}}. \par
The liquid viscosity is calculated using the Carreau model \cite{Carreau1972} with the power-law index 0.5 and 1.0 for shear-thinning and constant viscosity, respectively. It is calculated using,
\begin{alignat}{1}
	\mu_l(\dot{\gamma}) = \mu_o ( 1 + \left( \lambda \dot{\gamma}\right)^2 )^{\nicefrac{n - 1}{2}}
\end{alignat}
Where, $\mu_o$, $\lambda$ and $n$ are zero shear viscosity, time constant and power-law index, respectively. The shear rate $(\dot{\gamma})$ is given as,
\begin{alignat}{1}
	\dot{\gamma} = \sqrt{2\mathbf{D}:\mathbf{D}}
\end{alignat}
The deformation tensor $\mathbf{D}$ is defined as,
\begin{alignat}{1}
	\mathbf{D} = \nicefrac{1}{2}\left( \nabla \textbf{u} + \nabla \textbf{u}^{T} \right)
\end{alignat}
%
%
%
%
The dynamics of two bubbles ascending are governed by four non-dimensional parameters, 
\begin{alignat}{1}
	\text{E\"otv\"os number} &= Eo = \frac{\Delta \rho gd^2}{\sigma} \\
	\text{Morton number} &= Mo = \frac{g \mu_l^4 \Delta \rho}{\rho_l^2 \sigma^3} \\
	\text{Confinement ratio} &= C_r = \frac{W}{d} \\
	\text{center to center distance} &= h_d = \frac{h}{d}
\end{alignat}
\hspace{3mm} In the above equations, $\Delta \rho = \rho_l - \rho_g$, $W$ is width of liquid column and $h$ is the initial center to center distance of bubbles.  For a gas-liquid system the density ratio is large, hence $\Delta \rho = \rho_l - \rho_g \approx \rho_l$ and E\"otv\"os number can be also represented by Bond number. In the current study, the viscosity of the liquid phase $(\mu_l)$ varies with the shear rate $(\dot{\gamma})$, therefore we define the Morton number using the zero-shear liquid viscosity. Hence, the key non-dimensional parameters are given by: 
\begin{alignat}{1}
	\text{Bond number} &= Bo = \frac{\rho_l gd^2}{\sigma} \\
	\text{Morton number} &= Mo = \frac{g \mu_{l,0}^4}{\rho_l \sigma^3} \\
	\text{Confinement ratio} &= C_r = \frac{W}{d} \\
	\text{center to center distance} &= h_d = \frac{h}{d}
\end{alignat}

\section{Results and Discussion}
\label{sec:results}
We now present the results of a study to analyze the effects of fluid rheology on bubble deformation, rise velocity, rise path, and liquid viscosity distribution. Two power-law indices (0.5 and 1.0) have been considered to understand its effects. The power-law indices capture the shear-thinning and Newtonian behavior of the non-Newtonian fluid. The zero-shear viscosity $(\mu_0)$ and time constant $(\lambda)$ of the Carreau model \cite{Carreau1972} have been kept constant to isolate the effects of the power-law index. The bubbles are initially placed at the center of the cross-section, but the bubbles may come closer to the wall during the ascend. The Bond is kept constant at $Bo = 2$, and the terminal shape in the Newtonian fluid is expected to be ellipsoidal.\par 
%
%
%
Now, we discuss the effects of the power-law index $(n)$ for a center to center distance of $4d$. For two liquids considered here, the zero-shear Morton number $(Mo)$ is $10^{-6}$. According to the Grace diagram \cite{Grace1973}, an initially spherical bubble of $Bo = 2$ rising in a Newtonian fluid with a Morton number of $10^{-6}$ deforms into an ellipsoidal shape. In non-Newtonian fluids, the viscosity will be modified by the shear generated adjacent to the bubble. Hence, the effective viscosity will be lower for the shear-thinning fluid. Consequently, the effective Morton number of shear-thinning fluid will be lower than that of the Newtonian fluid. 
\begin{figure}[h]
	\begin{center}  
		\begin{tabular}{| *1{>{\centering\arraybackslash}m{0.2in}} | *1{>{\centering\arraybackslash}m{0.2in}} | *5{>{\centering\arraybackslash}m{0.375in}|} @{}m{0pt}@{}}
			\cline{3-7}
			\multicolumn{1}{c}{} & \multicolumn{1}{c|}{} & \multicolumn{5}{c|}{time (ms)} & \\ [2ex]
			\hline
			$n$ &  & \textbf{0} & \textbf{80} & \textbf{160}  & \textbf{240}  & \textbf{320} & \\[2ex] 
			\hline
			\textbf{0.5} & LB & 
			\includegraphics[width=0.375in,trim=4 4 4 4,clip]{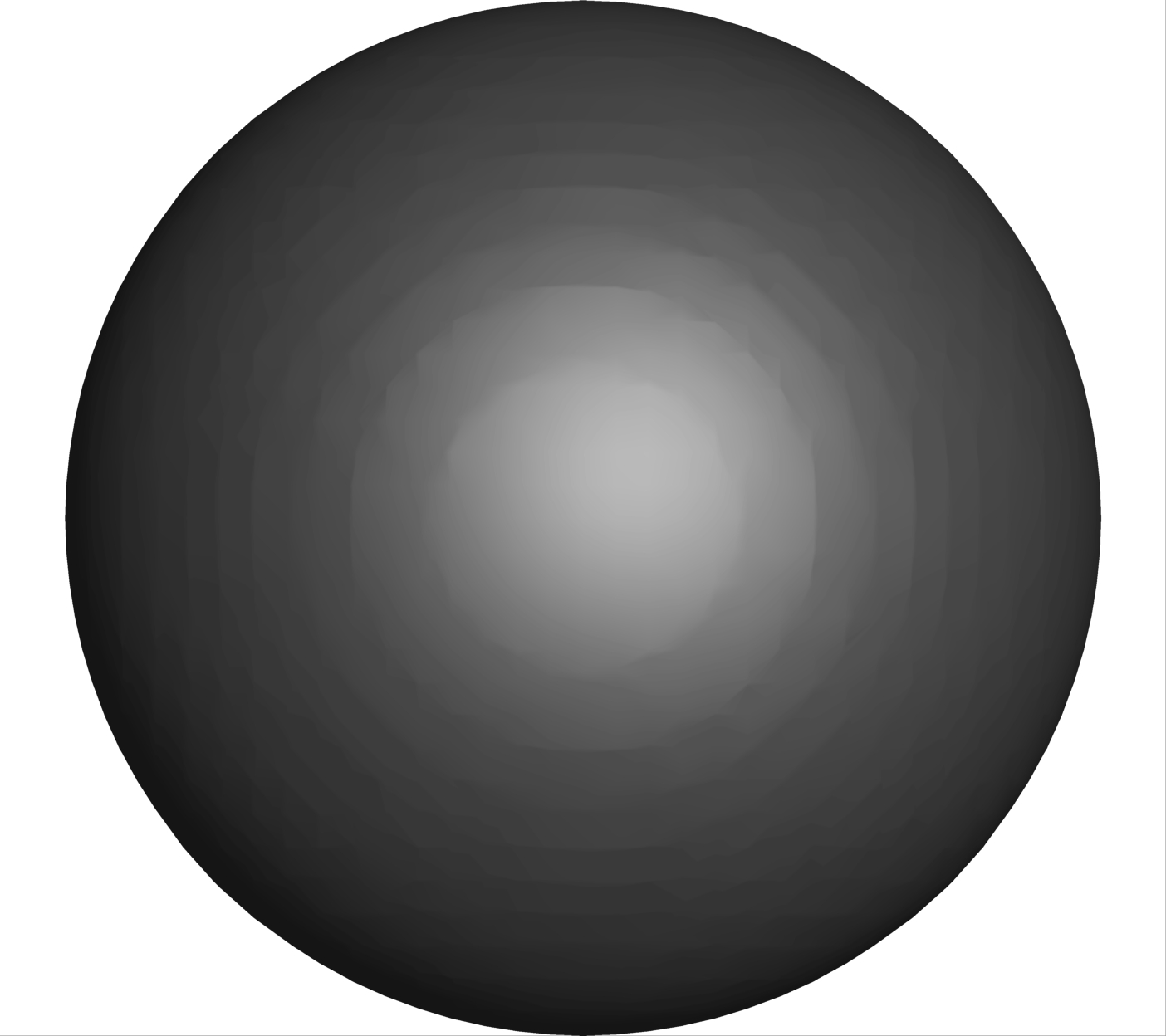} & \includegraphics[width=0.375in,trim=4 4 4 4,clip]{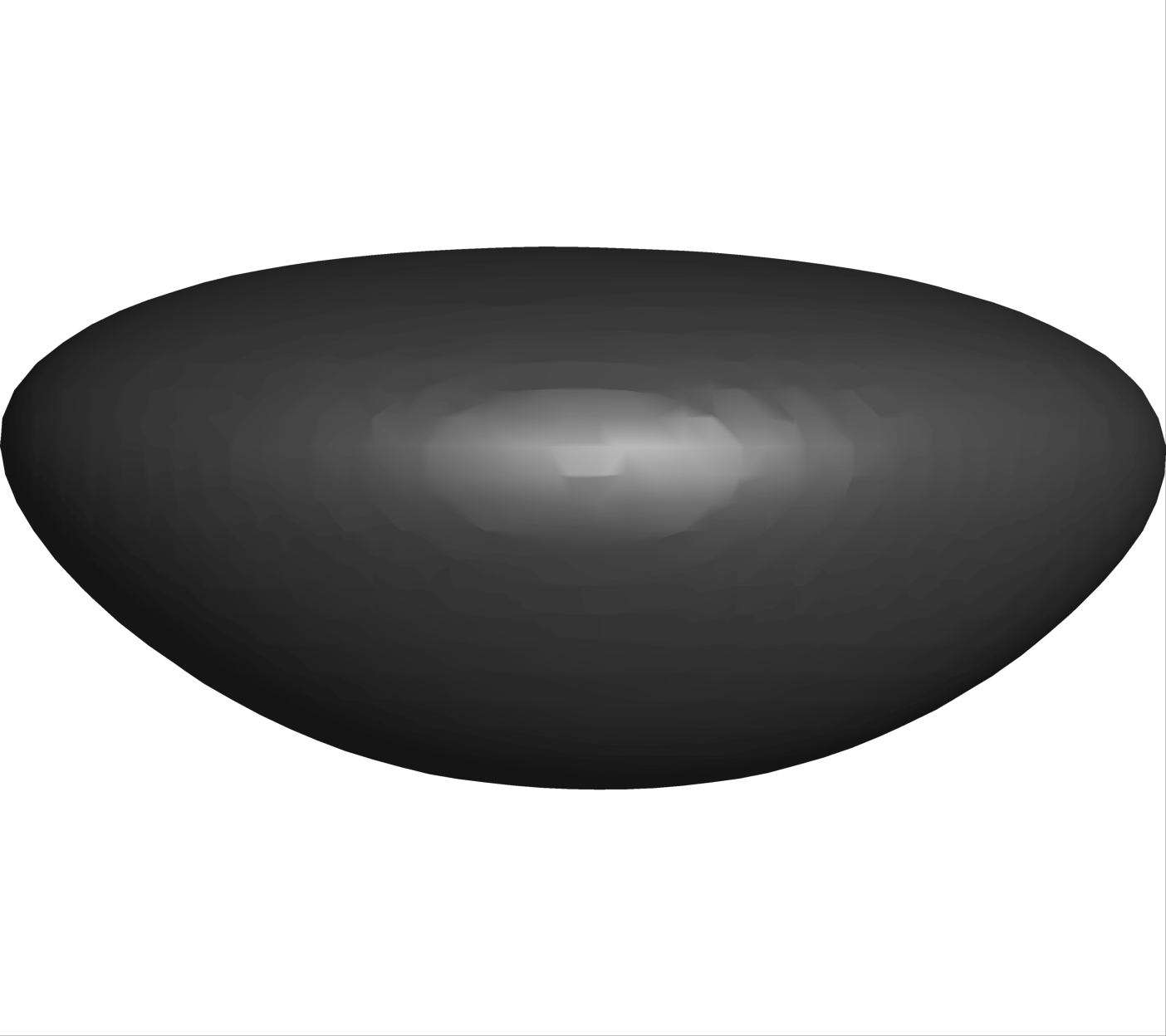} & \includegraphics[width=0.375in,trim=4 4 4 4,clip]{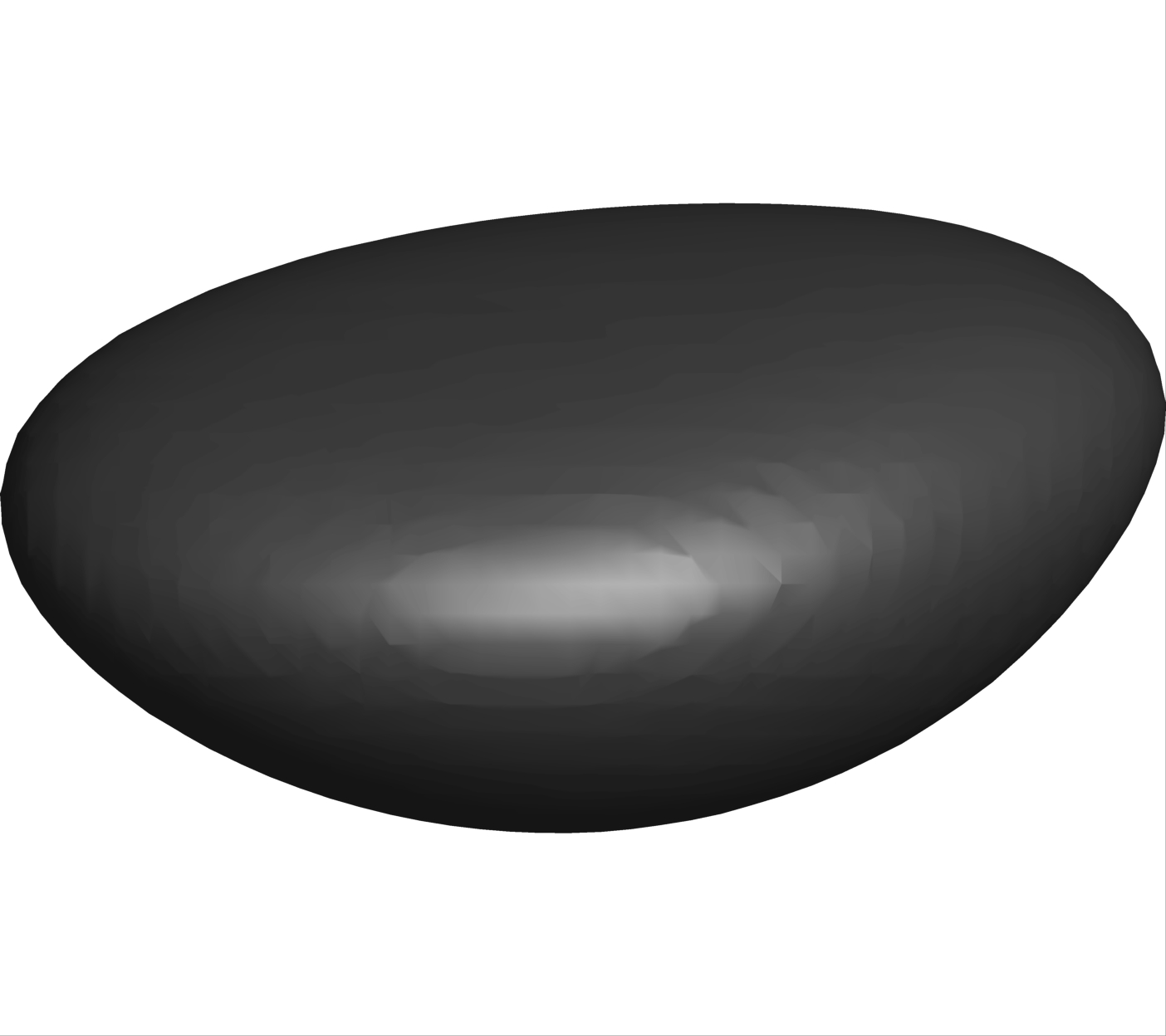} & \includegraphics[width=0.375in,trim=4 4 4 4,clip]{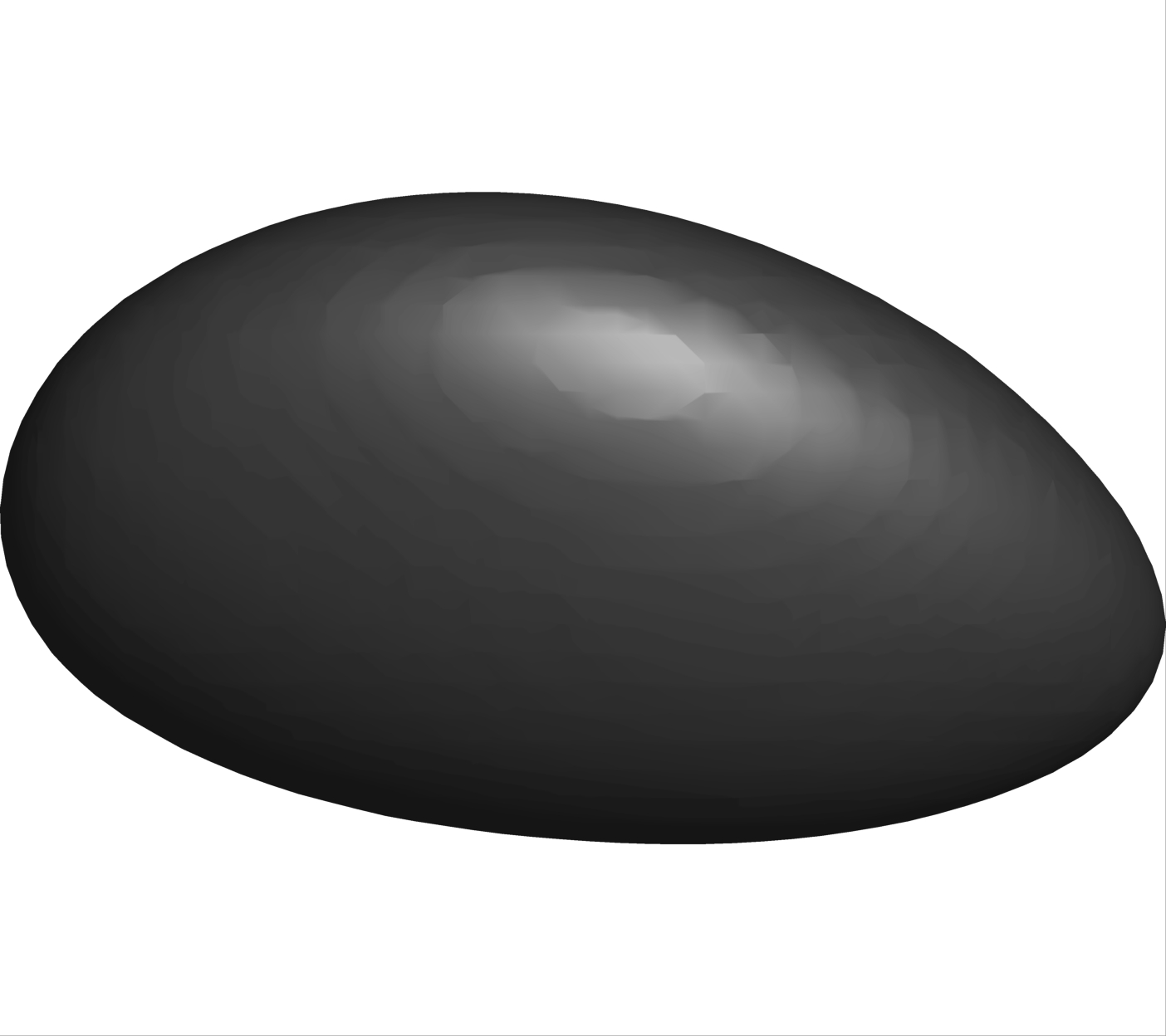} & \includegraphics[width=0.375in,trim=4 4 4 4,clip]{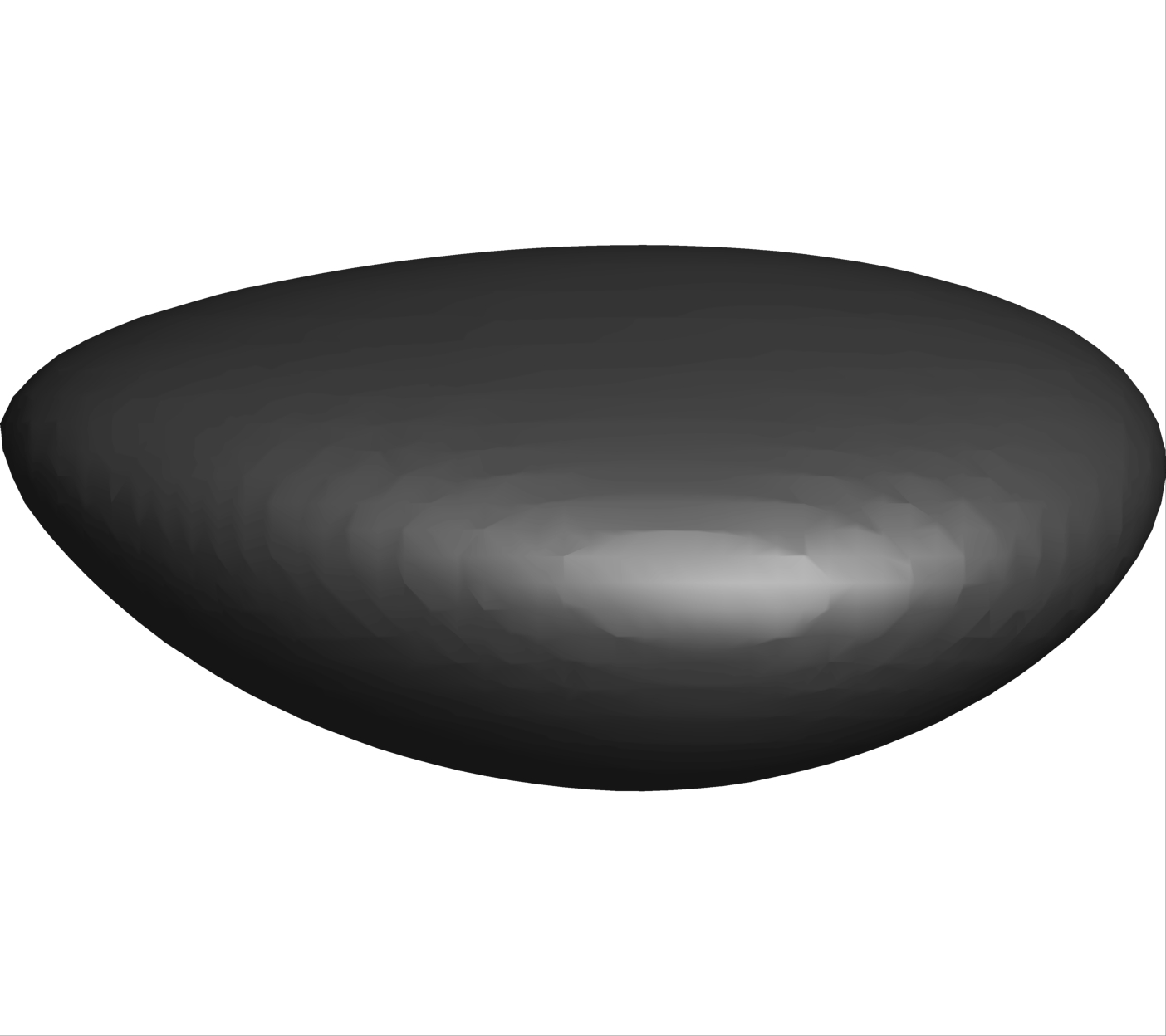} & \\[0ex]
			\hline
			\textbf{0.5} & TB &
			\includegraphics[width=0.375in,trim=4 4 4 4,clip]{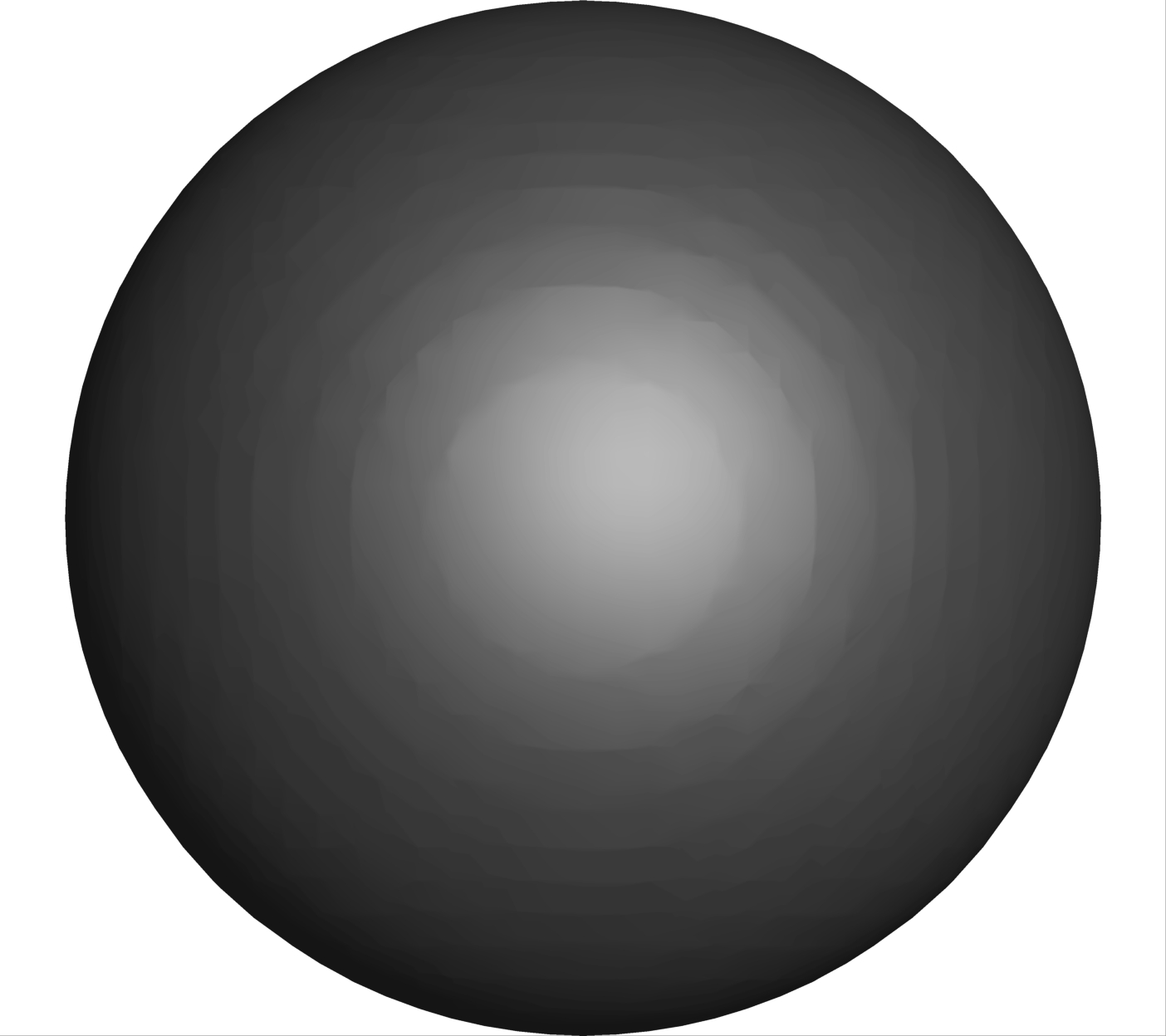} & \includegraphics[width=0.375in,trim=4 4 4 4,clip]{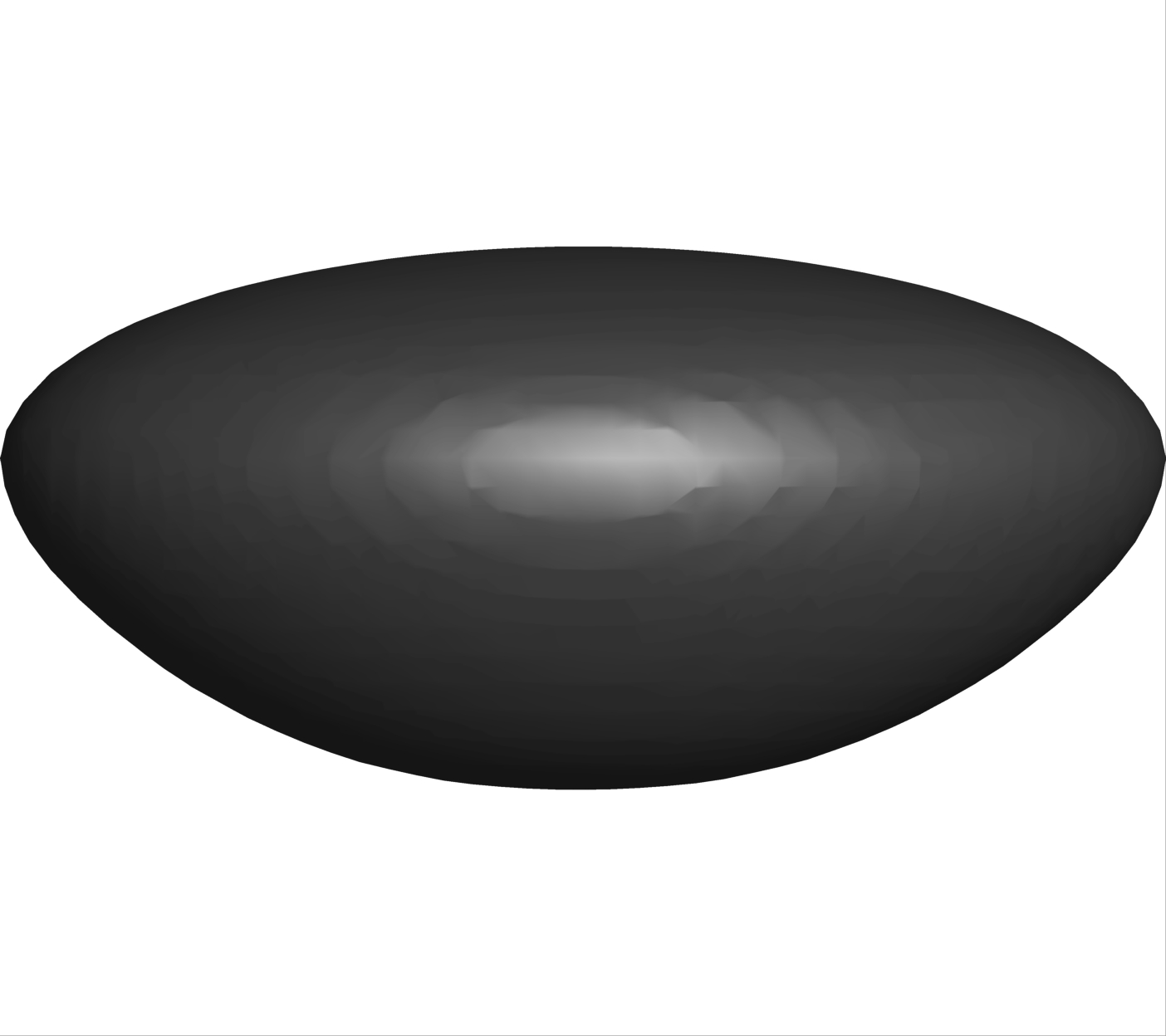} & \includegraphics[width=0.375in,trim=4 4 4 4,clip]{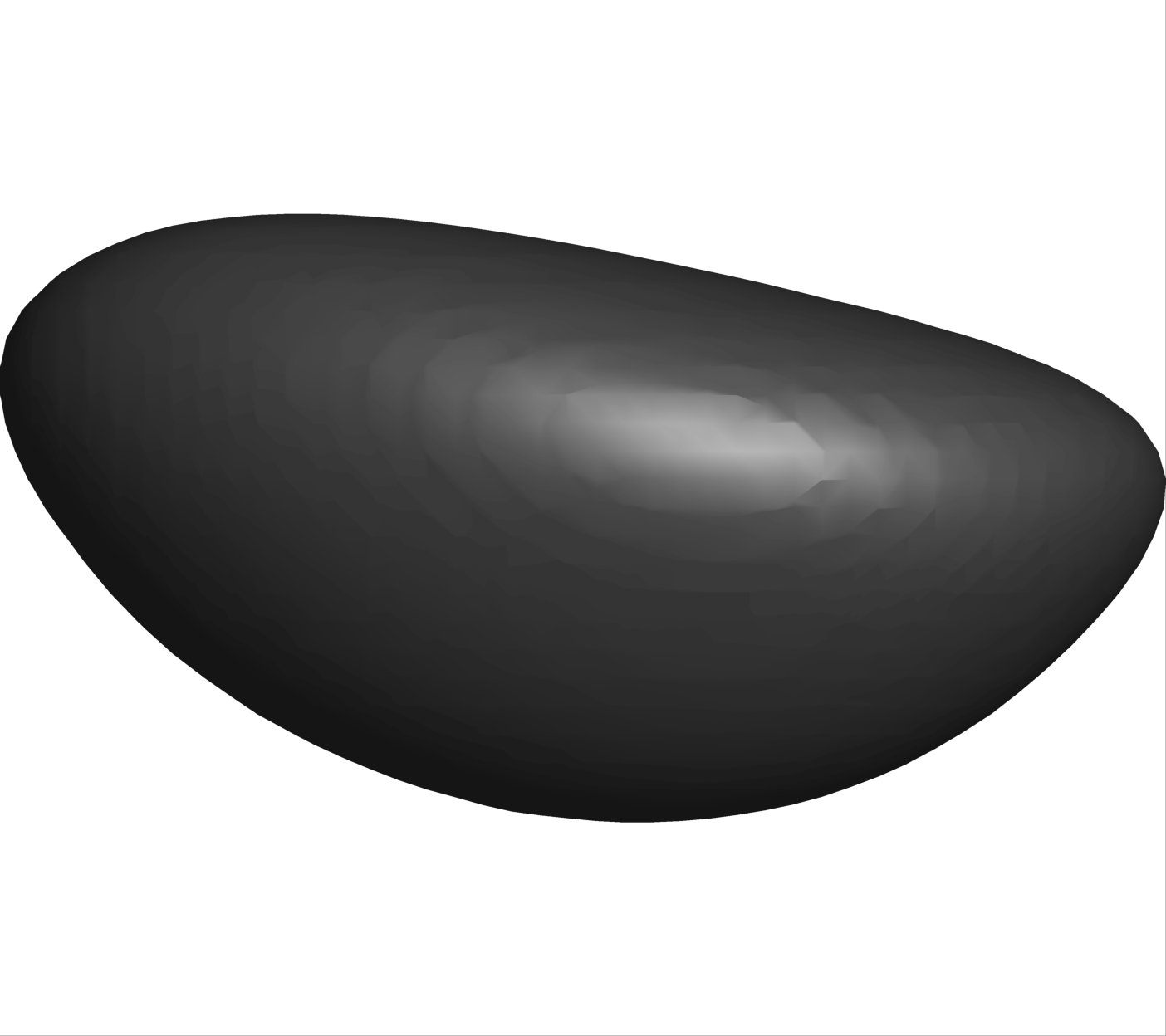} & \includegraphics[width=0.375in,trim=4 4 4 4,clip]{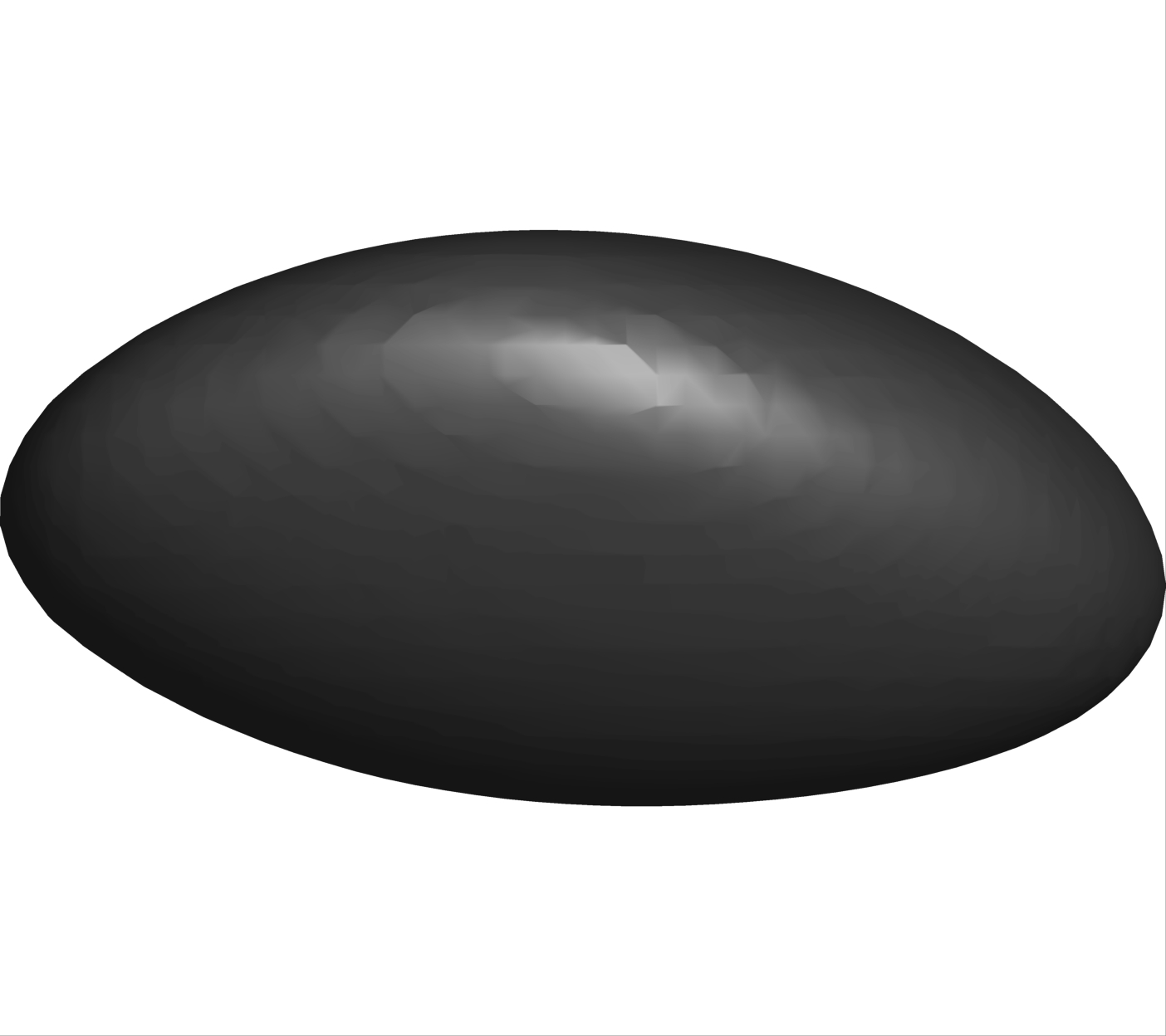} & \includegraphics[width=0.375in,trim=4 4 4 4,clip]{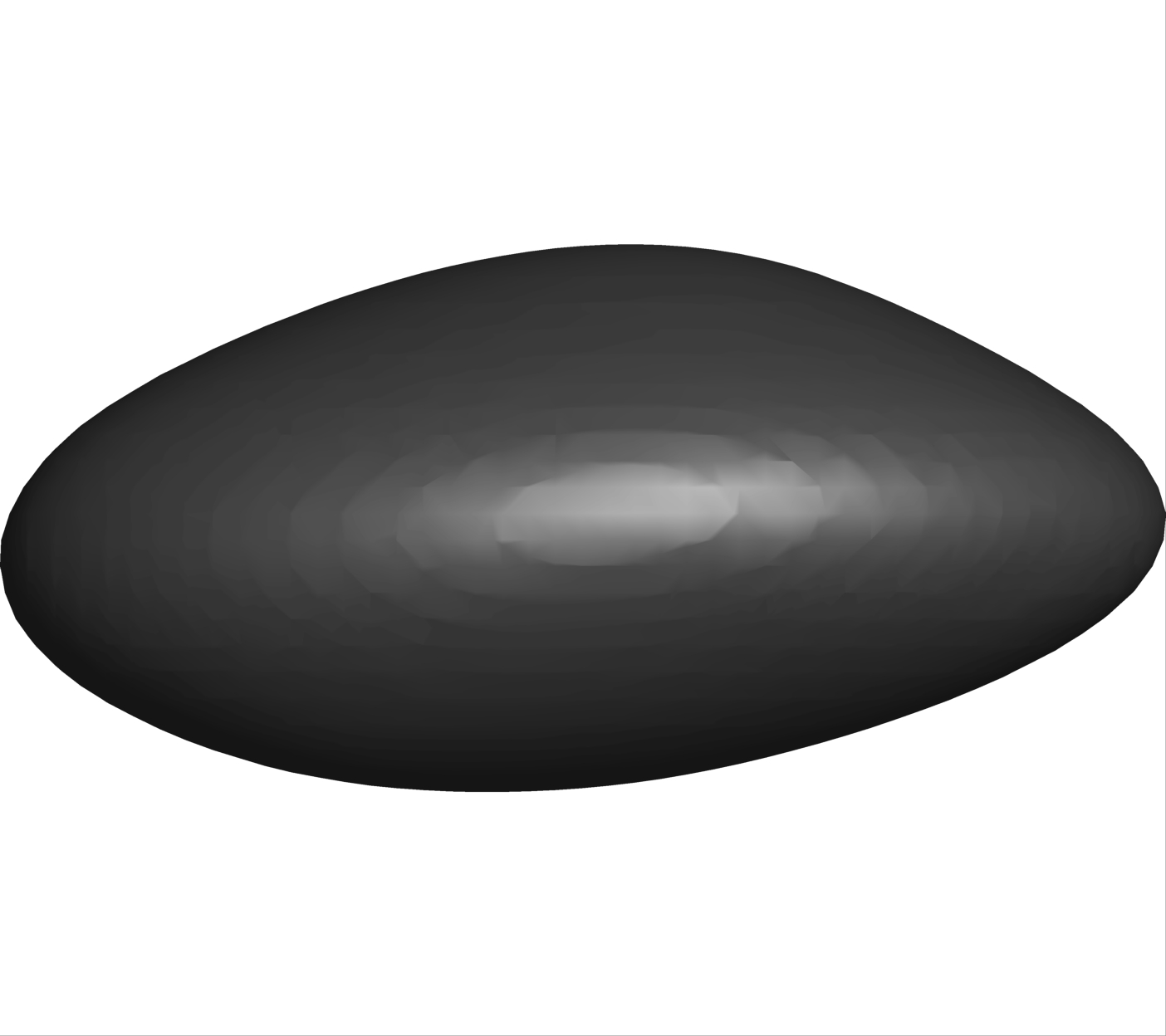} & \\[0ex]
			\hline
			\textbf{1.0} & LB &
			\includegraphics[width=0.375in,trim=4 4 4 4,clip]{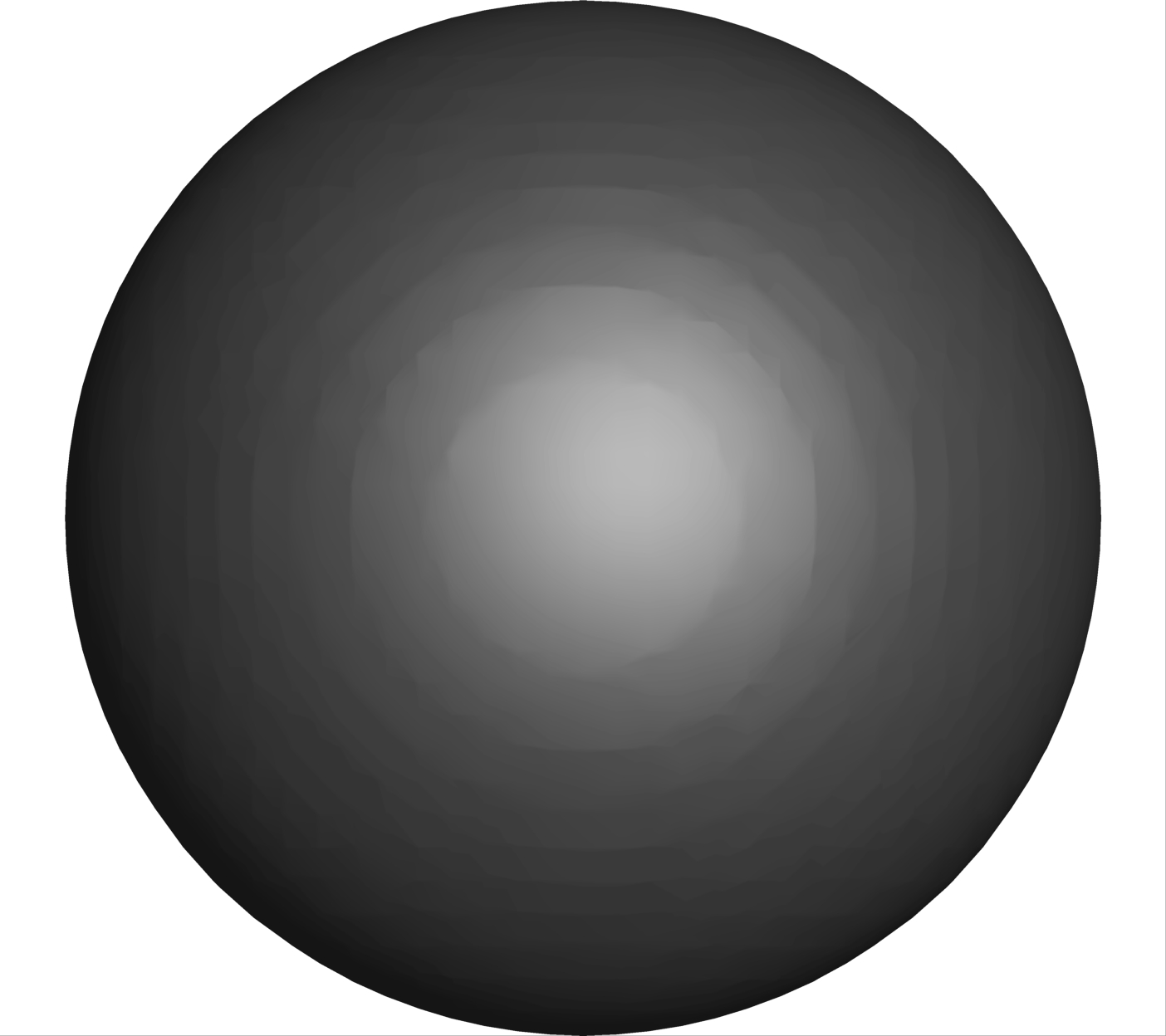} & \includegraphics[width=0.375in,trim=4 4 4 4,clip]{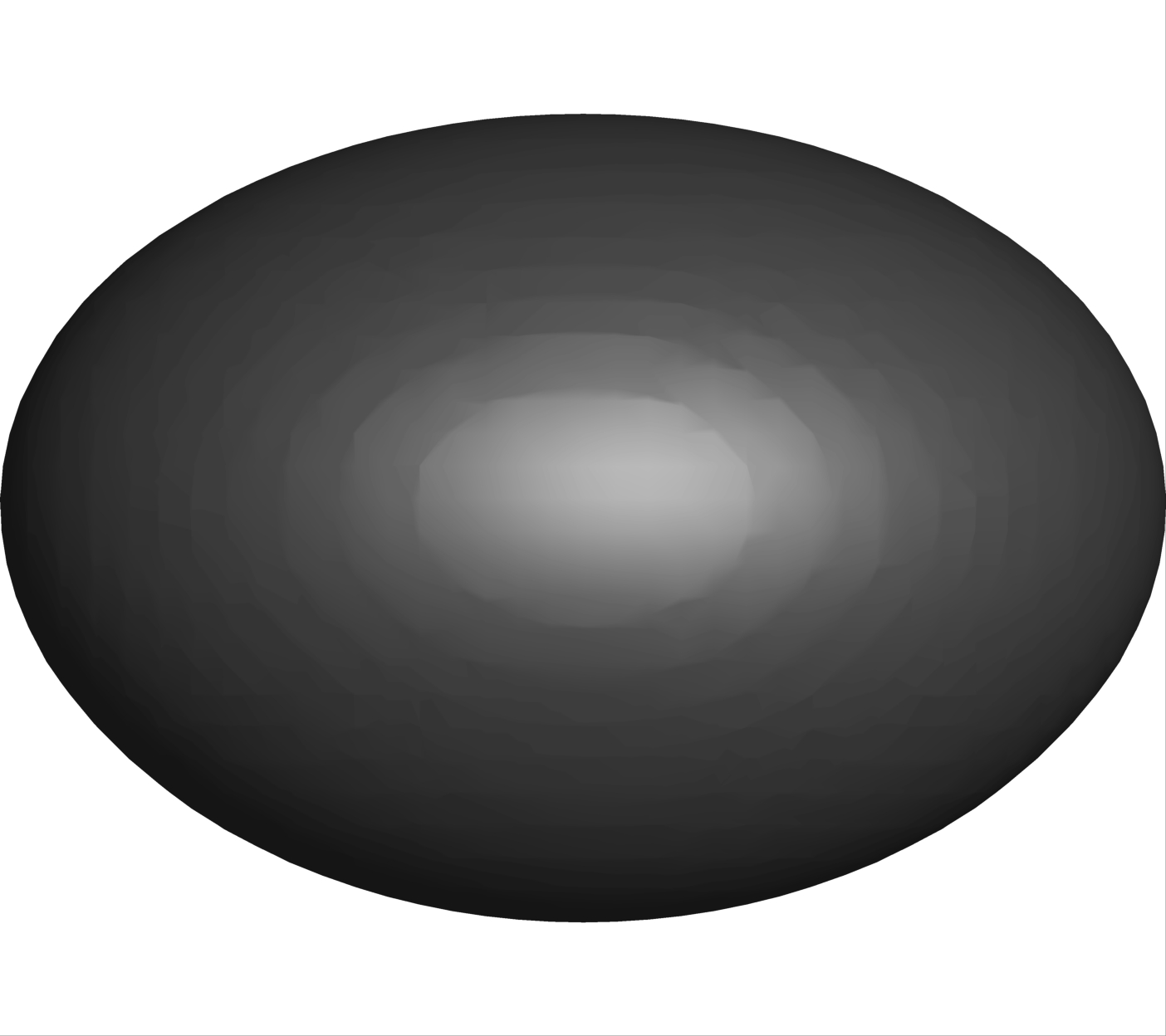} & \includegraphics[width=0.375in,trim=4 4 4 4,clip]{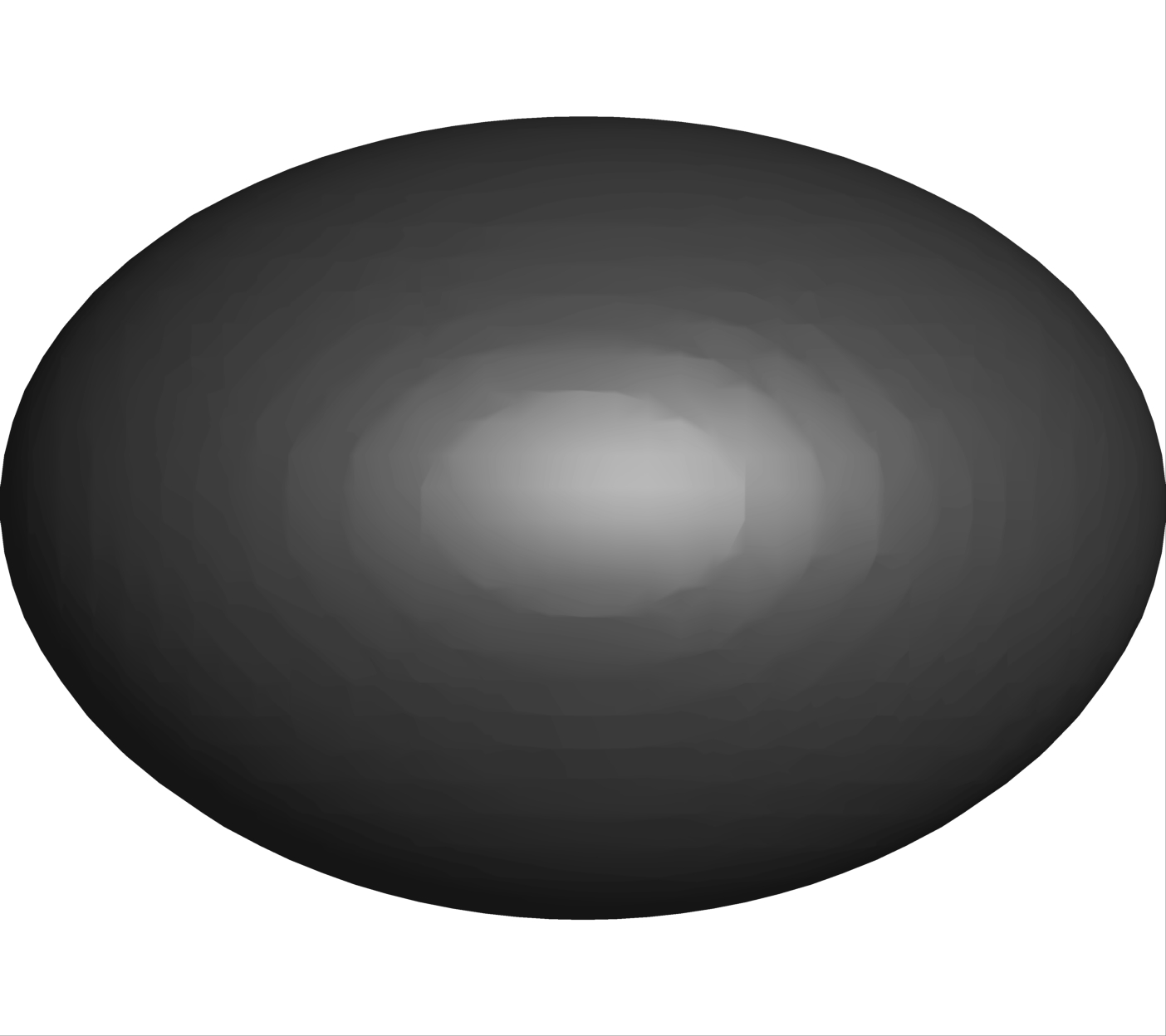} & \includegraphics[width=0.375in,trim=4 4 4 4,clip]{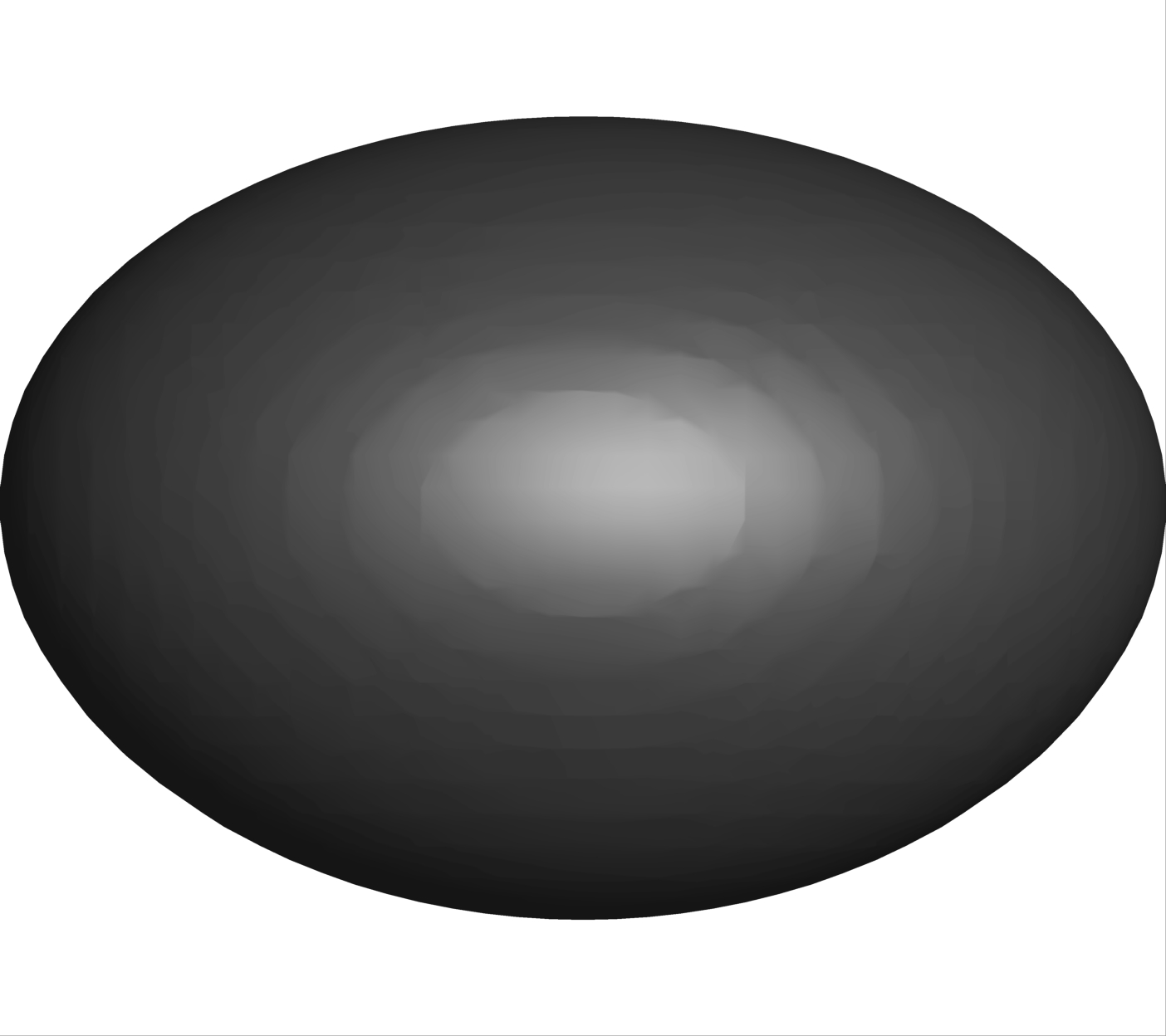} & \includegraphics[width=0.375in,trim=4 4 4 4,clip]{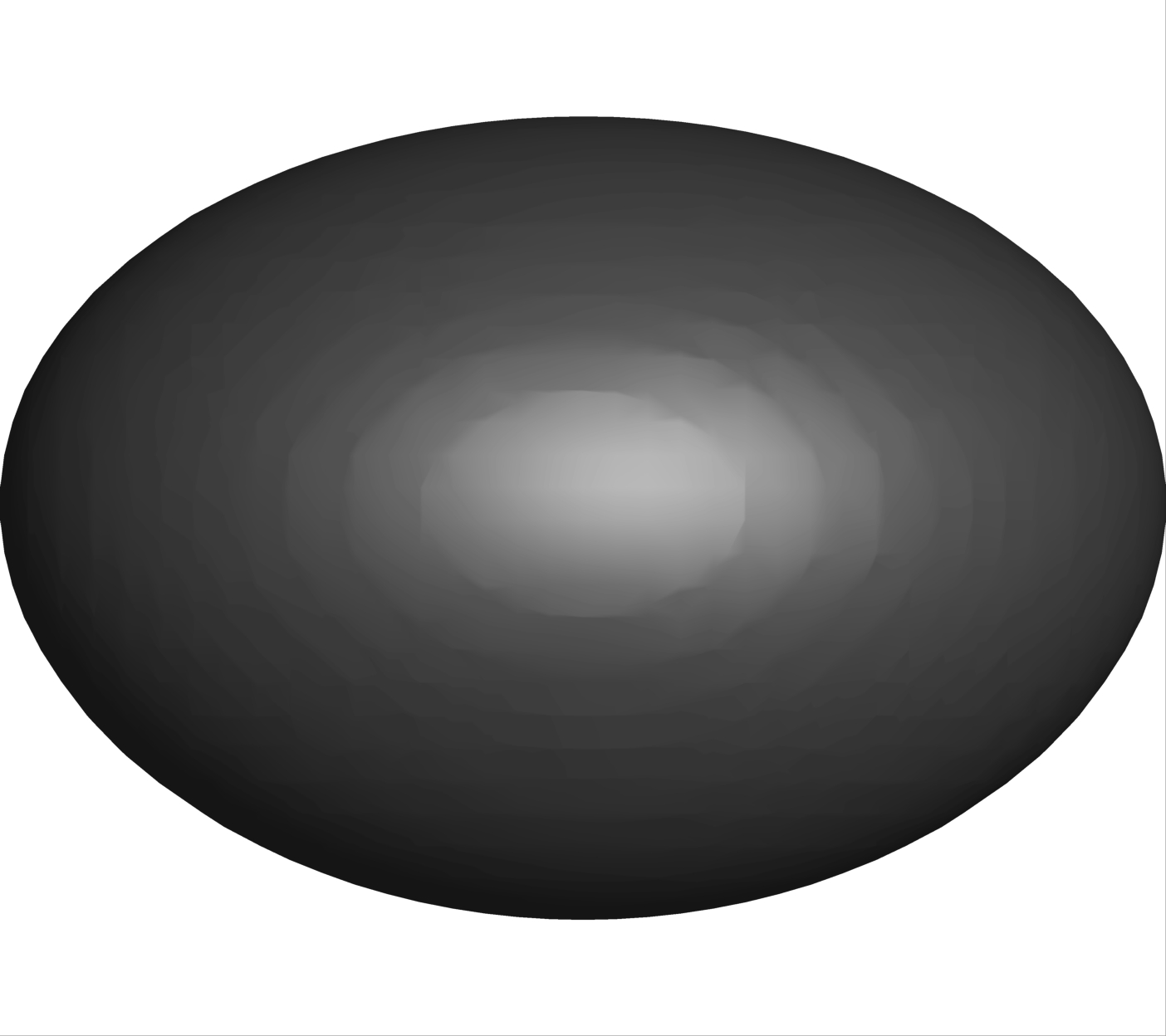} & \\[0ex]
			\hline
			\textbf{1.0} & TB &
			\includegraphics[width=0.375in,trim=4 4 4 4,clip]{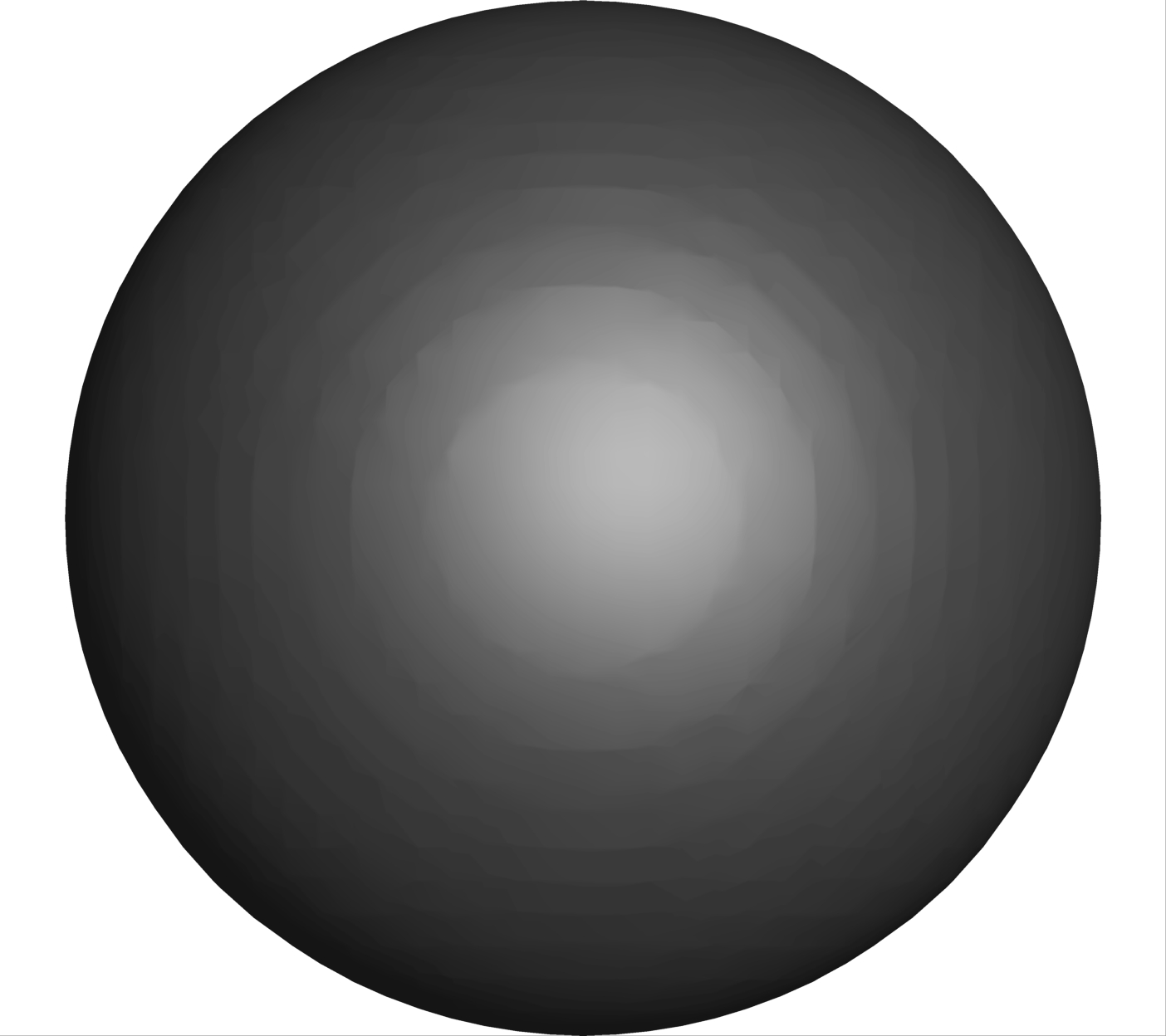} & \includegraphics[width=0.375in,trim=4 4 4 4,clip]{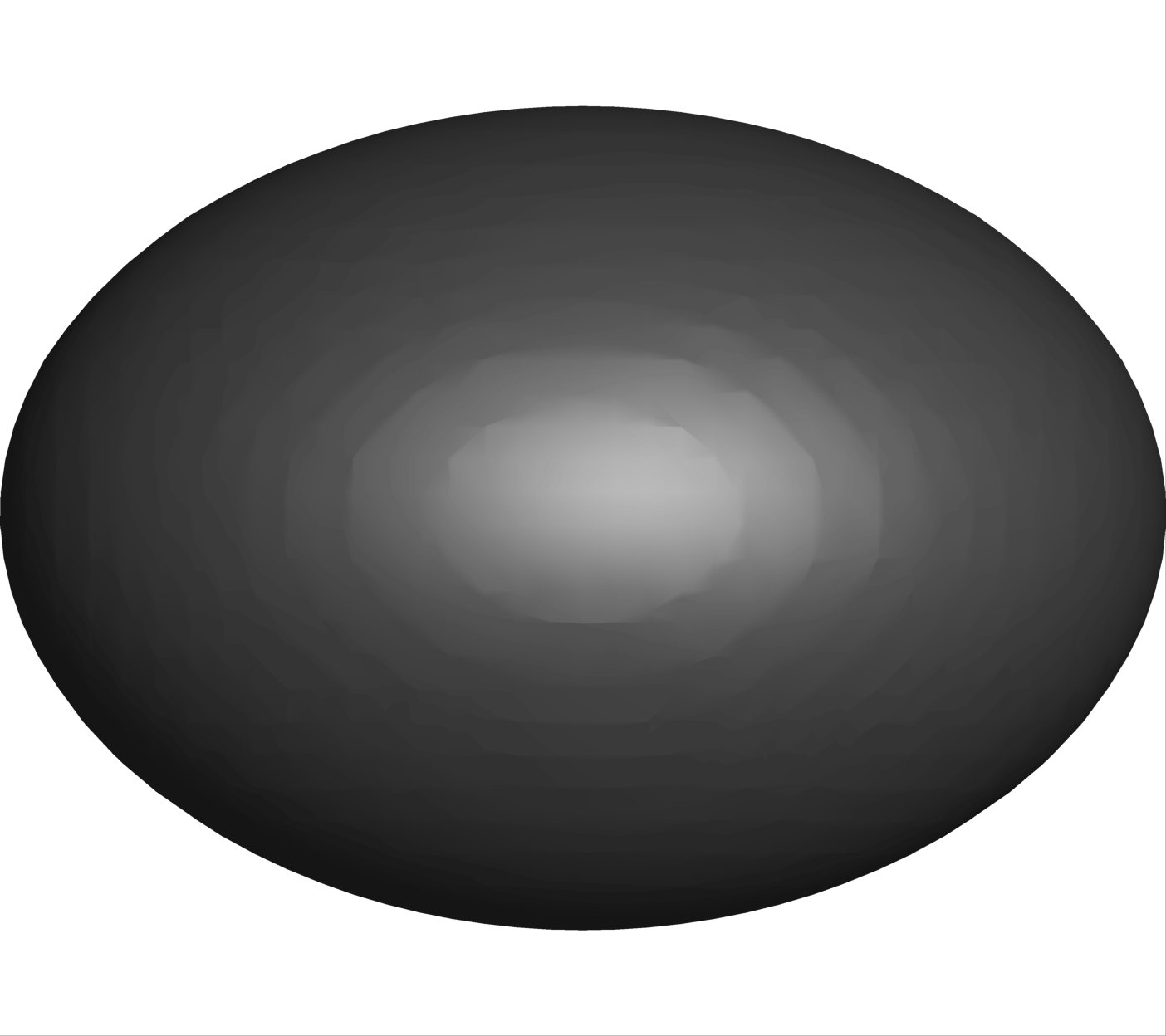} & \includegraphics[width=0.375in,trim=4 4 4 4,clip]{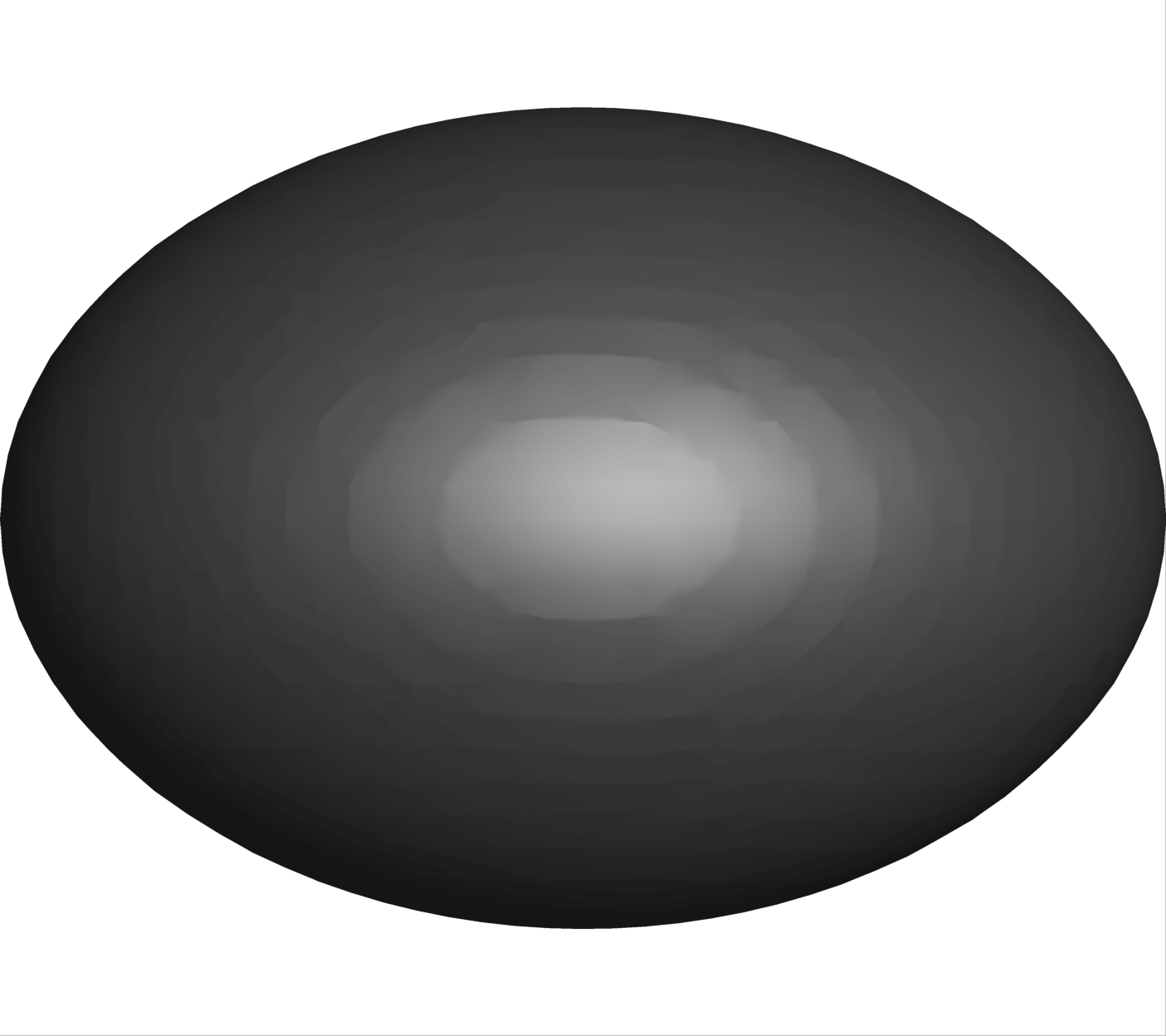} & \includegraphics[width=0.375in,trim=4 4 4 4,clip]{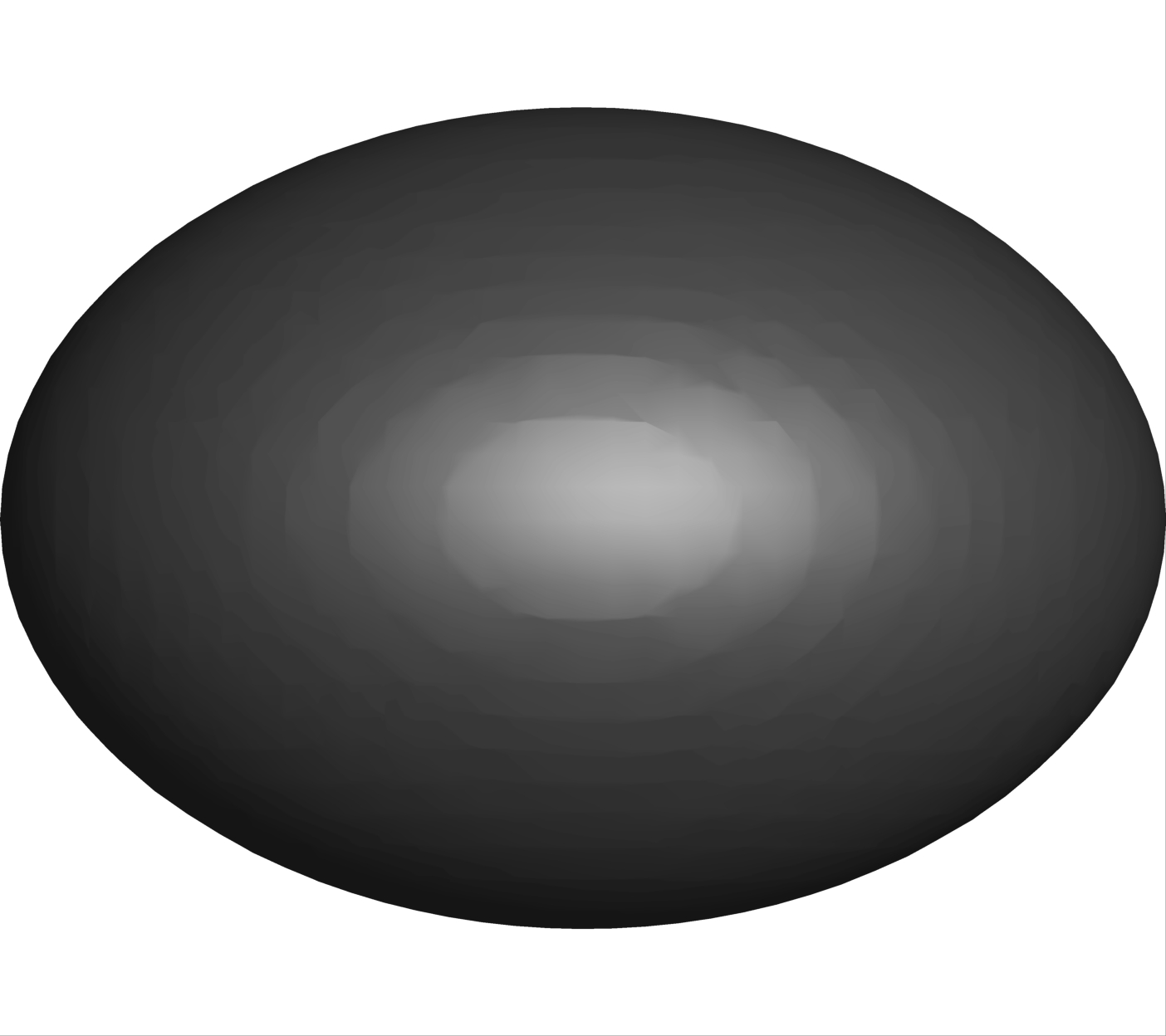} & \includegraphics[width=0.375in,trim=4 4 4 4,clip]{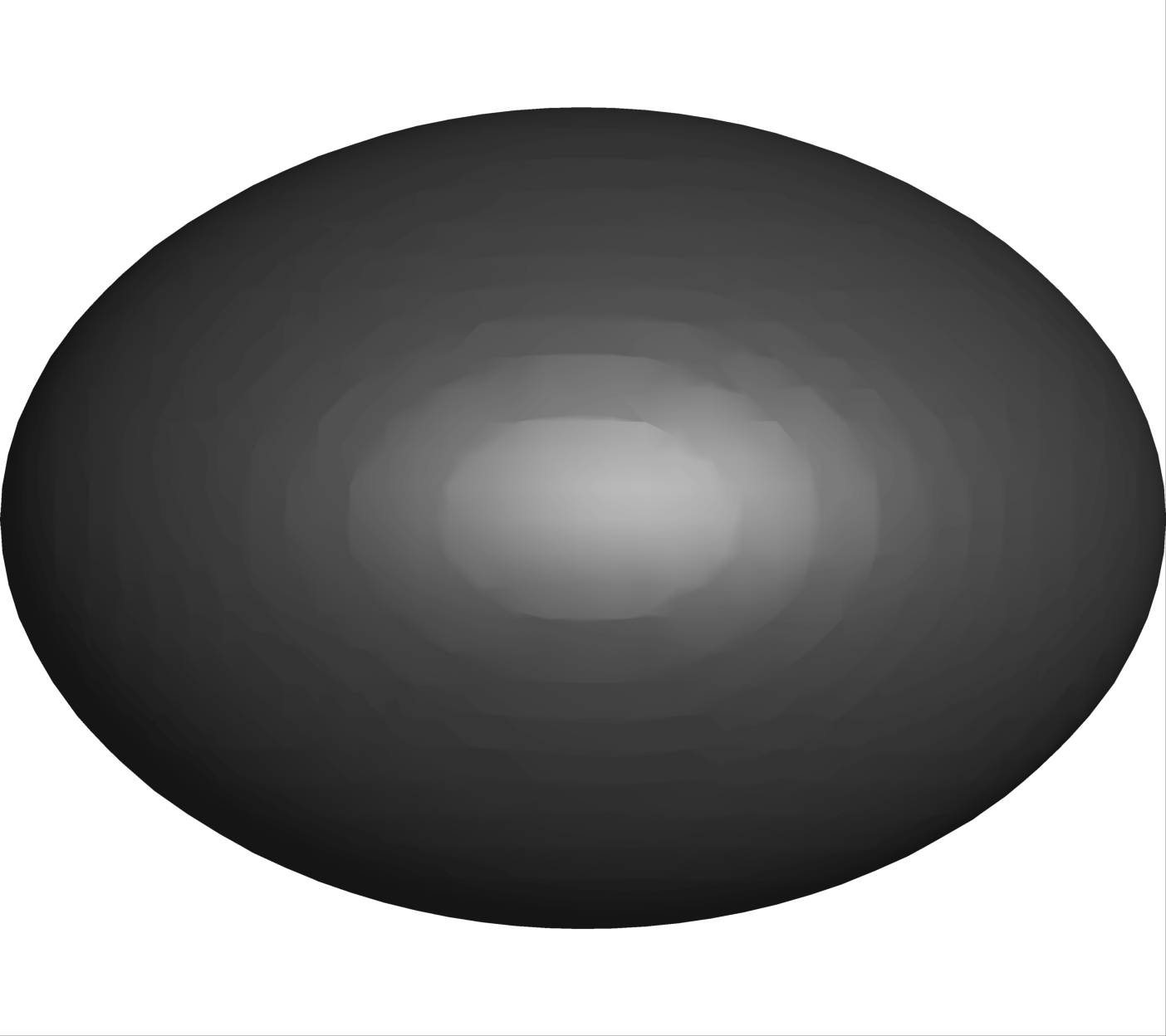} & \\[0ex]
			\hline
		\end{tabular}
		\caption{Transient three-dimensional front views of the bubbles for center to center distance = $4d$, and two power-law indices $(n)$}
		\label{fig:iso_surfaces_Mo_6_Bo_2_front}
	\end{center}
\end{figure}

Figure \ref{fig:iso_surfaces_Mo_6_Bo_2_front} shows the transient shapes of leading (LB) and trailing (TB) bubbles for $n$ = 0.5 and 1 and $h_d$ = 4. From fig. \ref{fig:iso_surfaces_Mo_6_Bo_2_front} we can observe that the bubble terminal shape in the Newtonian fluid (Fig. \ref{fig:iso_surfaces_Mo_6_Bo_2_front} row 3) is ellipsoidal with nearly symmetrical top and bottom surfaces. However, in the shear-thinning fluid (Fig. \ref{fig:iso_surfaces_Mo_6_Bo_2_front} row 1), the bubble deforms into an ellipsoid that is flatter than its Newtonian counterpart with definite asymmetry between top and bottom surfaces. It can also be noticed that the bubble in shear-thinning fluid goes through shape and orientation oscillations. In contrast, in the Newtonian fluid, the bubble achieves a steady shape and rises without any orientation changes. \par 
For the simulation time considered in this study, the dynamics of leading and trailing bubbles are quite similar in a Newtonian fluid. On the other hand, in shear-thinning fluid, their dynamics are complex. At $t = 80$ $ms$, the shape of both bubbles is nearly identical, primarily because the trailing bubble has not yet reached the flow field modified by the leading bubble. The orientation and shape of the leading and trailing bubbles are different afterward. At $t = 160$ $ms$, the surface normal at the top is upward-front and upward-back orientated for leading and trailing bubbles, respectively. This indicates that vortex shedding behind the bubble is oriented differently. At $t = 320$ $ms$, the shape of the trailing bubble is upside-down of the leading bubble.
\begin{figure}[h]
	\begin{center}
		\begin{subfigure}[b]{0.15\textwidth}
			\includegraphics[width=1\textwidth,trim=4 4 4 4,clip]{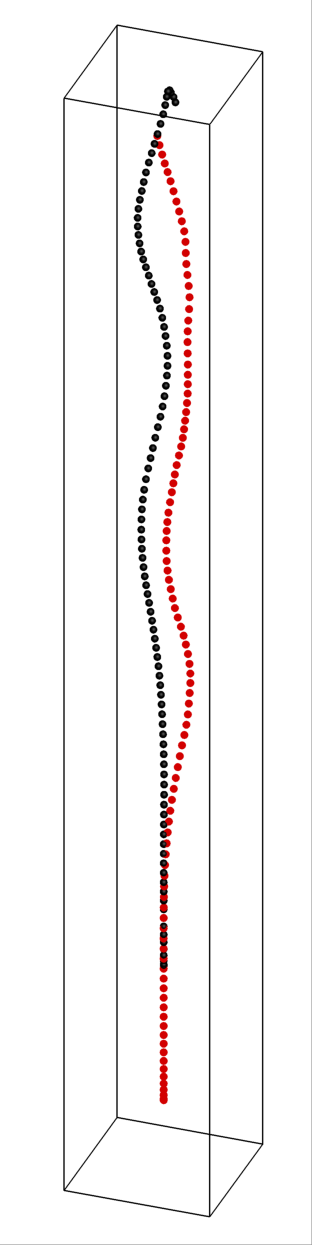}
			\caption{$n$ = 0.5}
			\label{fig:bubble_path_Bo_2_n_50_ivg_4d}
		\end{subfigure} %
		\begin{subfigure}[b]{0.15\textwidth}
			\includegraphics[width=1\textwidth,trim=4 4 4 4,clip]{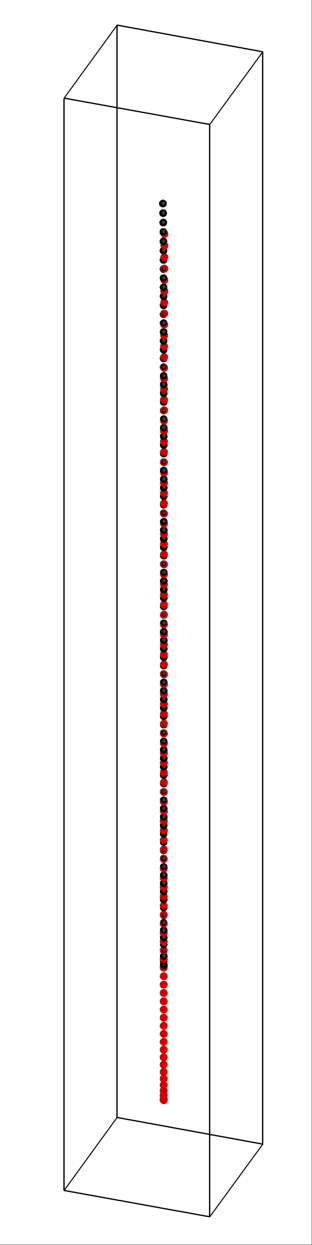}
			\caption{$n$ = 1.0}
			\label{fig:bubble_path_Bo_2_n_100_ivg_4d}
		\end{subfigure} %
		\caption{Bubble rise path for center to center distance = $4d$, and two power-law indices}
		\label{fig:bubble_path_hd_4d}
	\end{center}
\end{figure}

Figure \ref{fig:bubble_path_hd_4d} shows the ascending path of leading and trailing bubbles for a center to center distance of $4d$ and two power-law indices. It can be noticed that bubbles in the Newtonian fluid rise in a rectilinear path. This is in line with our observation from Fig. \ref{fig:iso_surfaces_Mo_6_Bo_2_front}, the bubble reached a steady shape without any shape or orientation oscillations. It has also been reported earlier by Kumar et al. [???] that in the absence of bubble shape or orientation oscillations, the vortices behind the bubble are symmetric and do not amplify any instability. We are also observing similar phenomena for both bubbles in the Newtonian fluid.

In the shear-thinning fluid, the rise path of bubbles goes from rectilinear to oscillatory. It should be noted that initially the trailing bubble has a smaller velocity and is away from the flow field modified by the leading bubble. Its shape changes from the spherical to asymmetric ellipsoid and doesn't exhibit surface oscillations. This is the primary reason for the initial rectilinear path of the bubble. As the bubble accelerates, it reaches a certain critical speed (or Reynold number), after which it sheds asymmetric vortices. This leads to non-uniform pressure distribution behind the bubble, which pushes the bubble away from the center of the liquid column and creates a zig-zag rise path. From Fig. \ref{fig:bubble_path_Bo_2_n_50_ivg_4d} we can observe that the leading bubble initially migrates towards the left compared to the trailing bubble migrating towards the right.
%
\begin{figure}[h]
	\begin{center}
		\includegraphics[width=0.45\textwidth,trim=0 0 0 0,clip]{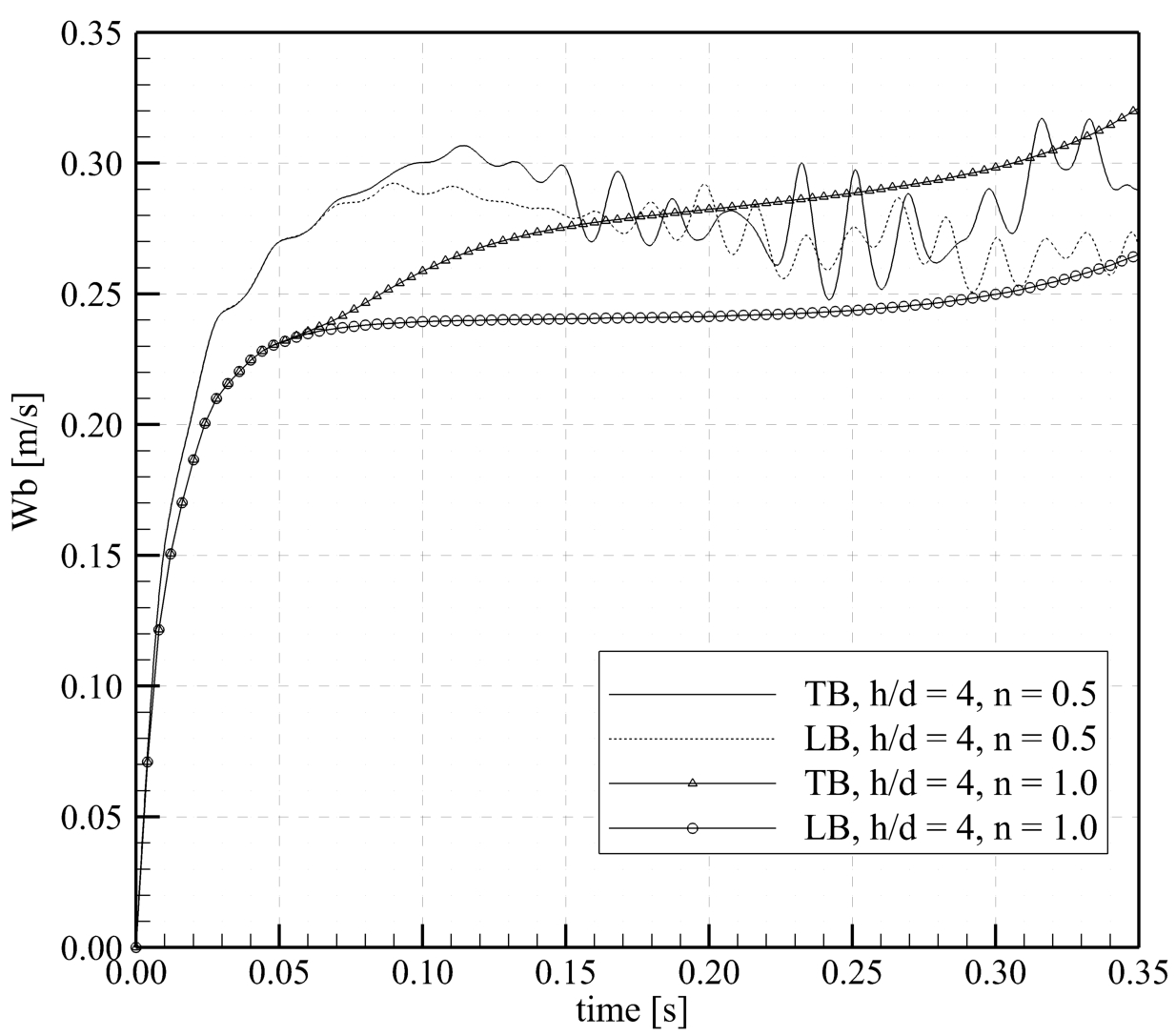}
		\caption{Rise velocity of bubbles for center to center distance = $4d$, and two power-law indices}
		\label{fig:rise_vel_hd_4d}
	\end{center}
\end{figure}

Figure \ref{fig:rise_vel_hd_4d} shows the rise velocities of leading and trailing bubbles in the Newtonian and shear-thinning fluids for a center to center distance of 4. In the Newtonian fluid, the velocity curves of both bubbles are identical for up to $t \approx 60 ~ms$, and afterward, the curves diverge. The velocity of the leading bubble between $60~ms$ and $200~ms$ is almost constant, whereas the trailing bubble rapidly accelerates to achieve a higher velocity than the leading bubble. This behavior is primarily because the leading bubble ascends through the undisturbed flow field hence experiences maximum drag. The trailing bubble ascends through a modified flow field; as a result, it experiences lower drag force and achieves higher rise velocity.

Similar to the Newtonian case, rise velocity curves in the shear-thinning fluid are identical for up to $t \approx 70 ~ms$, and afterward, the curves diverge. The rise velocity goes through rapid oscillations. In the Newtonian fluid, the trailing bubble velocity was always higher than that of the leading bubble. However, in the shear-thinning fluid, there are instances where the leading bubble rises with higher velocity than the trailing bubble. It can be noted for any given time window, the amplitude of oscillation of trailing bubble rise velocity is higher than that of the leading bubble. \par

\begin{figure}[h]
	\begin{center}
		\begin{subfigure}[b]{0.23\textwidth}
			\includegraphics[width=1\textwidth,trim=4 4 4 4,clip]{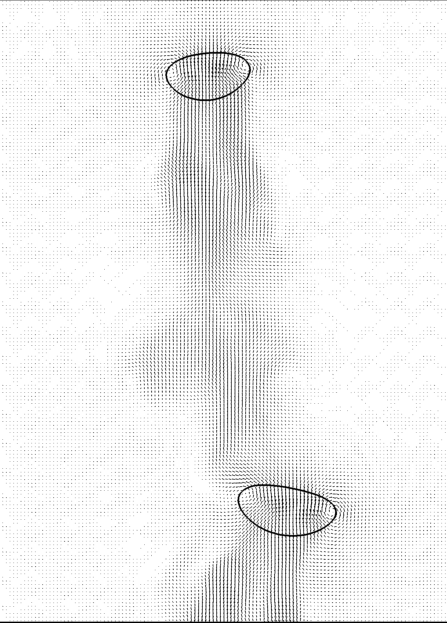}
			\caption{$n = 0.5$}
			\label{fig:velvector_hd_4d_n_0p5}
		\end{subfigure} %
		\hspace{1mm}
		\begin{subfigure}[b]{0.23\textwidth}
			\includegraphics[width=1\textwidth,trim=4 4 4 4,clip]{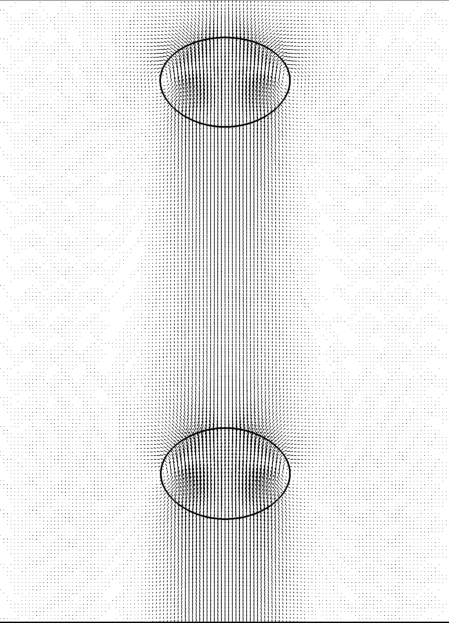}
			\caption{$n = 1.0$}
			\label{fig:velvector_hd_4d_n_1p0}
		\end{subfigure} %
		\caption{Velocity vector at $t = 160~ms$ for center to center distance = $4d$, and two power-law indices }
		\label{fig:vel_vec_hd_4d}
	\end{center}
\end{figure}

Figures \ref{fig:velvector_hd_4d_n_0p5} and \ref{fig:velvector_hd_4d_n_1p0} show the velocity vectors on $yz-$plane at $t = 160~ms$ for shear-thinning and constant viscosity cases. The interface has been drawn to show the cross-section of the bubble on the $yz-$ plane. We notice from Fig. \ref{fig:velvector_hd_4d_n_1p0} that velocity vectors in front of the trailing bubble are streamlined without any recirculation zones, and on both left and right sides, the flow distribution is symmetric. From Fig. \ref{fig:velvector_hd_4d_n_0p5} we can notice that the flow field in front of the trailing bubble contains recirculation zones and is asymmetric on the left and right sides of the bubble. The interaction between the recirculation zone and bubble creates asymmetric forces and pushes away from the centerline. The impact of this was observed in Fig. \ref{fig:bubble_path_Bo_2_n_50_ivg_4d}.

\begin{figure}[h]
	\begin{center}
		\begin{subfigure}[b]{0.23\textwidth}
			\includegraphics[width=1\textwidth,trim=4 4 4 4,clip]{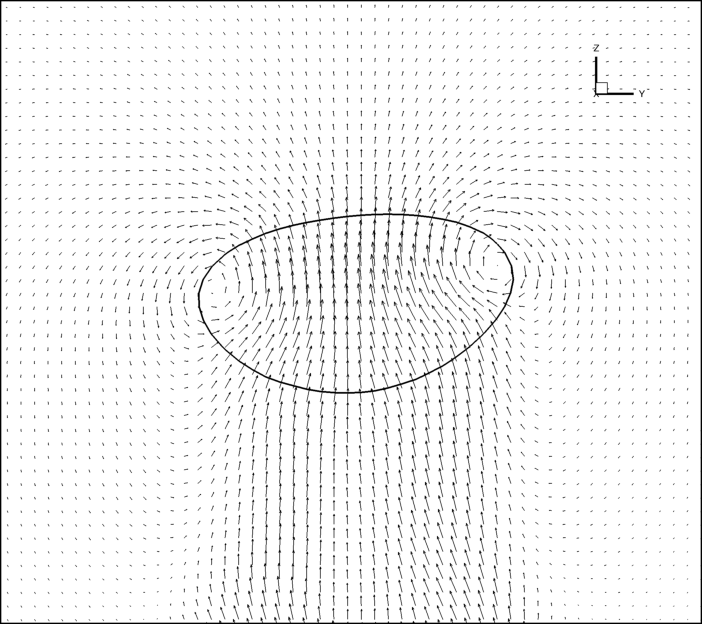}
			\caption{Leading bubble}
			\label{fig:velvector_hd_4d_n_0p5_LB}
		\end{subfigure} %
		\begin{subfigure}[b]{0.23\textwidth}
			\includegraphics[width=1\textwidth,trim=4 4 4 4,clip]{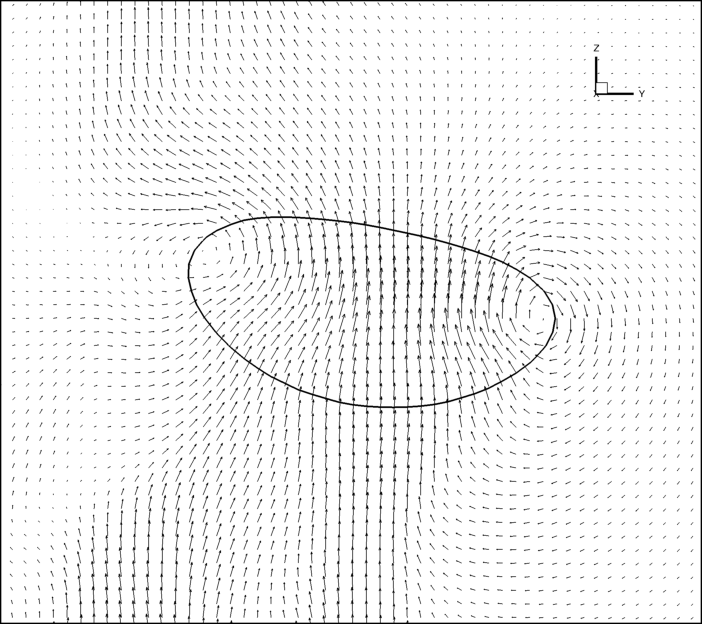}
			\caption{Trailing bubble}
			\label{fig:velvector_hd_4d_n_0p5_TB}
		\end{subfigure} %
		\caption{Velocity vector at $t = 160~ms$ for center to center distance = $4d$, and two power-law index = 0.5 }
		\label{fig:vel_vec_hd_4d_zoomed_n_0p5}
	\end{center}
\end{figure}

\begin{figure}[h]
	\begin{center}
		\begin{subfigure}[b]{0.23\textwidth}
			\includegraphics[width=1\textwidth,trim=4 4 4 4,clip]{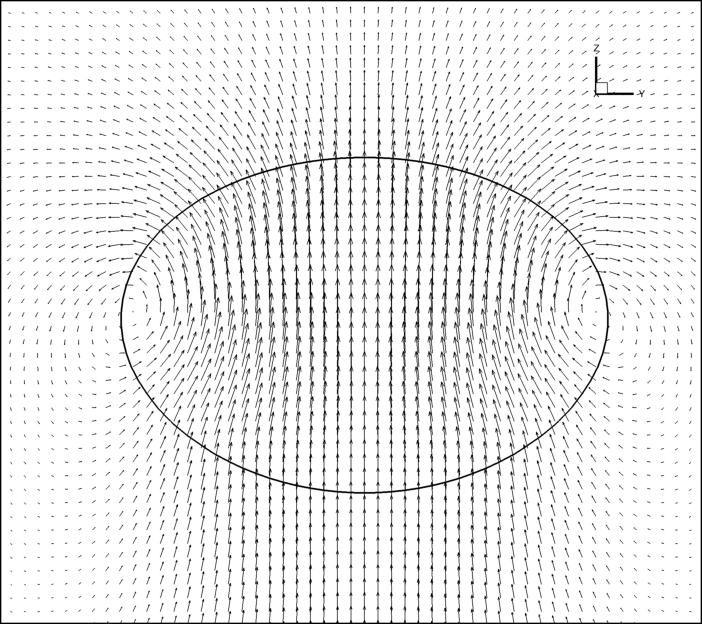}
			\caption{Leading bubble}
			\label{fig:velvector_hd_4d_n_1p0_LB}
		\end{subfigure} %
		\begin{subfigure}[b]{0.23\textwidth}
			\includegraphics[width=1\textwidth,trim=4 4 4 4,clip]{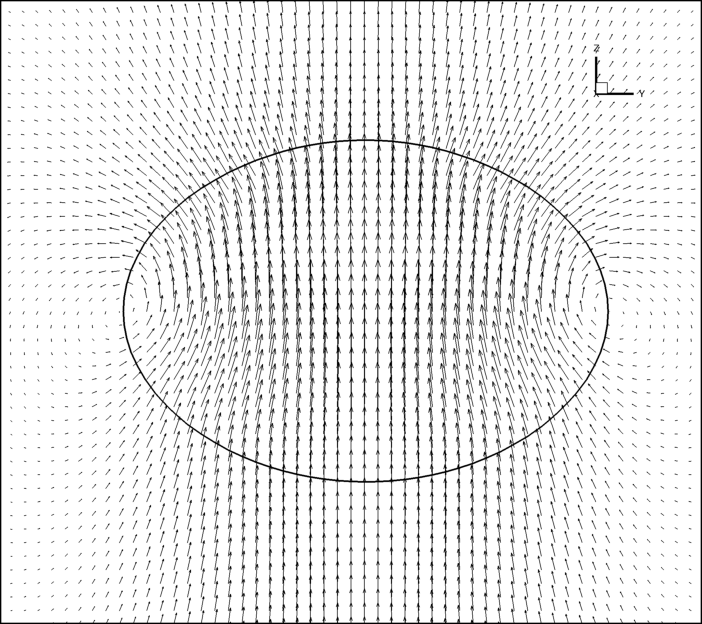}
			\caption{Trailing bubble}
			\label{fig:velvector_hd_4d_n_1p0_TB}
		\end{subfigure} %
		\caption{Velocity vector at $t = 160~ms$ for center to center distance = $4d$, and two power-law index = 1.0 }
		\label{fig:vel_vec_hd_4d_zoomed_n_1p0}
	\end{center}
\end{figure}

Figures \ref{fig:vel_vec_hd_4d_zoomed_n_0p5} and \ref{fig:vel_vec_hd_4d_zoomed_n_1p0} show the zoomed-in view of bubbles at $t = 160~ms$ on the $yz-$plane. The zoom ratio of the four subfigures is kept the same to show that the deformation of the bubble in the shear-thinning fluid is substantially higher than that in the Newtonian fluid. We can further observe the asymmetric and symmetric flow field around the trailing bubbles in shear-thinning and Newtonian fluid, respectively.

\section{Conclusions}
\label{sec:conclusions}
In this numerical study, we have compared the effects of fluid rheology on the three-dimensional dynamics of two in-line bubbles rising in a quiescent liquid medium. We have used a volume of fluid method coupled with a three-dimensional geometry construction method for accurate representation and advection of the interface. We have modeled the surface tension using the sharp surface force (SSF) method. The fractional step method with the implicit treatment of the viscous terms and geometry construction method for the convection term has been used to solve the Navier-Stokes equations. The effects of the power-law index on the bubble shapes, rise velocity, rise path, and flow field at representative times have been explored. \par
We observe that the bubble deformation is strongly dependent on the power-law index of the fluid -- in the Newtonian fluid, both leading and trailing bubbles deformed into an ellipsoid with symmetric top and bottom surfaces and attained steady-state shape. In contrast, the bubble became more flatter with asymmetric top and bottom surfaces in the shear-thinning fluid and went through surface oscillations as they ascended.  

Further, we observe that the bubbles initially rose in a linear path in shear-thinning fluid and later followed a zigzag path. In contrast, in the Newtonian fluid, they rose in a rectilinear path throughout the ascent. The zigzag motion of the leading bubble was triggered by an asymmetry in the flow field behind and consequent vortex shedding behind the bubble. The zigzag motion of the trailing bubble is the combination of the unsteady flow in front of it and the asymmetric flow around it.

The rise velocity oscillates in the shear-thinning fluids, whereas no such features are present in the Newtonian and shear-thickening fluids. These oscillations are related to the shape oscillation of the bubble and a zigzag rise path. In both liquids, leading and trailing bubbles initially ascended with identical speeds. Once they achieved maximum deformation, the trailing bubble in the Newtonian fluid rose faster than the leading bubble. In the shear-thinning fluid, the magnitude of oscillation of rising velocity was higher for the trailing bubble than the leading bubble.  

%
%

\bibliographystyle{unsrtnat}	
\bibliography{ms.bib}

\begin{thebibliography}{79}
\providecommand{\natexlab}[1]{#1}
\providecommand{\url}[1]{\texttt{#1}}
\expandafter\ifx\csname urlstyle\endcsname\relax
  \providecommand{\doi}[1]{doi: #1}\else
  \providecommand{\doi}{doi: \begingroup \urlstyle{rm}\Url}\fi

\bibitem[Morsi and Alexander(1972)]{Morsi1972particle}
S.~A. Morsi and A.~J. Alexander.
\newblock An investigation of particle trajectories in two-phase flow systems.
\newblock \emph{Journal of Fluid Mechanics}, 55:\penalty0 193--208, 9 1972.
\newblock ISSN 1469-7645.
\newblock \doi{10.1017/S0022112072001806}.

\bibitem[Kurose et~al.(2001)Kurose, Misumi, and Komori]{Kurose2001drag_lift}
R.~Kurose, R.~Misumi, and S.~Komori.
\newblock Drag and lift forces acting on a spherical bubble in a linear shear
  flow.
\newblock \emph{International Journal of Multiphase Flow}, 27\penalty0
  (7):\penalty0 1247 -- 1258, 2001.
\newblock ISSN 0301-9322.
\newblock \doi{http://dx.doi.org/10.1016/S0301-9322(00)00073-2}.

\bibitem[Magnaudet and Legendre(1998)]{Magnaudet1998drag}
Jacques Magnaudet and Dominique Legendre.
\newblock The viscous drag force on a spherical bubble with a time-dependent
  radius.
\newblock \emph{Physics of Fluids}, 10\penalty0 (3), 1998.

\bibitem[Adoua et~al.(2009)Adoua, Legendre, and
  Magnaudet]{ADOUA2009Lift_force_reversal}
Richard Adoua, Dominique Legendre, and Jacques Magnaudet.
\newblock Reversal of the lift force on an oblate bubble in a weakly viscous
  linear shear flow.
\newblock \emph{Journal of Fluid Mechanics}, 628:\penalty0 23--41, 6 2009.
\newblock ISSN 1469-7645.
\newblock \doi{10.1017/S0022112009006090}.
\newblock URL \url{http://journals.cambridge.org/article_S0022112009006090}.

\bibitem[Allen(1900)]{Allen1900}
H.~S. Allen.
\newblock The motion of a sphere in a viscous fluid.
\newblock \emph{Philosophical Magazine Series 5}, 50\penalty0 (304):\penalty0
  323--338, 1900.

\bibitem[Hadamard(1911)]{Hadamard1911}
J.~Hadamard.
\newblock Mouvement permanent lent d'une sphere liquide et visqueuse dans une
  liquide visqueux.
\newblock \emph{Comptes rendus de l'Académie des sciences}, 152:\penalty0
  1735, 1911.

\bibitem[Rybczynski(1911)]{Rybczynski1911}
W.~Rybczynski.
\newblock On the translatory motion of a fluid sphere in a viscous medium.
\newblock \emph{Bull. Acad. Sci., Krakow, Series A}, 40:\penalty0 40--46, 1911.

\bibitem[Bond(1927)]{Bond1927}
W.N. Bond.
\newblock Bubbles and drops and {Stokes'} law.
\newblock \emph{Philosophical Magazine Series 7}, 4\penalty0 (24):\penalty0
  889--898, 1927.

\bibitem[Moore(1959)]{Moore1959}
D.~W. Moore.
\newblock {The rise of a gas bubble in a viscous liquid}.
\newblock \emph{Journal of Fluid Mechanics}, 6\penalty0 (01):\penalty0
  113--130, March 1959.
\newblock ISSN 0022-1120.

\bibitem[Kumar et~al.(2015{\natexlab{a}})Kumar, Jin, and
  Vanka]{Kumar2015non_newtonian}
Purushotam Kumar, Kai Jin, and S.~Pratap Vanka.
\newblock Bubble rise and deformation in a non-newtonian fluid in a square
  duct.
\newblock \emph{Computational Methods and Tools in Thermal Fluids Sciences},
  pages 1--15, 2015{\natexlab{a}}.

\bibitem[Grace(1973)]{Grace1973}
JR~Grace.
\newblock Shapes and velocities of bubbles rising in infinite liquids.
\newblock \emph{Trans. Inst. Chem. Eng.}, 51\penalty0 (2):\penalty0 116--120,
  1973.

\bibitem[Grace et~al.(1976)Grace, Wairegi, and Nguyen]{Grace1976}
J.R. Grace, T.~Wairegi, and T.H Nguyen.
\newblock Shapes and velocities of single drops and bubbles moving freely
  through immiscible liquids.
\newblock \emph{Chemical Engineering Research and Design}, 54a:\penalty0
  167--173, 1976.

\bibitem[Bhaga and Weber(1981)]{Bhaga1981}
Dahya Bhaga and M.E. Weber.
\newblock Bubbles in viscous liquids: shapes, wakes and velocities.
\newblock \emph{Journal of Fluid Mechanics}, 105:\penalty0 61--85, 1981.

\bibitem[Figueroa-Espinoza et~al.(2008)Figueroa-Espinoza, Zenit, and
  Legendre]{Figueroa-Espinoza2008}
B.~Figueroa-Espinoza, R.~Zenit, and D.~Legendre.
\newblock {The effect of confinement on the motion of a single clean bubble}.
\newblock \emph{Journal of Fluid Mechanics}, 616:\penalty0 419--443, November
  2008.
\newblock ISSN 0022-1120.

\bibitem[Kumar and Vanka(2015)]{Kumar2015confinement}
P.~Kumar and S.P. Vanka.
\newblock Effects of confinement on bubble dynamics in a square duct.
\newblock \emph{International Journal of Multiphase Flow}, 77:\penalty0 32 --
  47, 2015.
\newblock ISSN 0301-9322.

\bibitem[Clift et~al.(1978)Clift, Grace, and Weber]{Clift1978}
Roland Clift, John~R Grace, and Martin~E Weber.
\newblock \emph{Bubbles, Drops and Particles}.
\newblock Academic Press, London, 1978.

\bibitem[DeKee(2002)]{DeKee2002}
Daniel DeKee.
\newblock \emph{Transport Processes in Bubbles, Drops and Particles}.
\newblock CRC Press, 2002.

\bibitem[Astarita and Apuzzo(1965)]{Astarita1965}
Gianni Astarita and Gennaro Apuzzo.
\newblock Motion of gas bubbles in non-newtonian liquids.
\newblock \emph{AIChE Journal}, 11\penalty0 (5):\penalty0 815--820, 1965.
\newblock ISSN 1547-5905.
\newblock \doi{10.1002/aic.690110514}.
\newblock URL \url{http://dx.doi.org/10.1002/aic.690110514}.

\bibitem[Leal et~al.(1971)Leal, Skoog, and Acrivos]{Leal1971}
L.~G. Leal, J.~Skoog, and A.~Acrivos.
\newblock On the motion of gas bubbles in a viscoelastic liquid.
\newblock \emph{The Canadian Journal of Chemical Engineering}, 49\penalty0
  (5):\penalty0 569--575, 1971.
\newblock ISSN 1939-019X.
\newblock \doi{10.1002/cjce.5450490504}.
\newblock URL \url{http://dx.doi.org/10.1002/cjce.5450490504}.

\bibitem[Carreau et~al.(1974)Carreau, Devic, and Kapellas]{Carreau1974}
P.~J. Carreau, M.~Devic, and M.~Kapellas.
\newblock Dynamique des bulles en milieu visco{\'e}lastique.
\newblock \emph{Rheologica Acta}, 1974.

\bibitem[Zana and Leal(1978)]{Zana1978}
E.~Zana and L.G. Leal.
\newblock The dynamics and dissolution of gas bubbles in a viscoelastic fluid.
\newblock \emph{International Journal of Multiphase Flow}, 4\penalty0
  (3):\penalty0 237 -- 262, 1978.
\newblock ISSN 0301-9322.
\newblock \doi{http://dx.doi.org/10.1016/0301-9322(78)90001-0}.

\bibitem[Kawase and Ulbrecht(1981)]{Kawase1981}
Yoshinori Kawase and Jaromir~J. Ulbrecht.
\newblock On the abrupt change of velocity of bubbles rising in non-newtonian
  liquids.
\newblock \emph{Journal of Non-Newtonian Fluid Mechanics}, 8\penalty0
  (3–4):\penalty0 203 -- 212, 1981.
\newblock ISSN 0377-0257.
\newblock \doi{http://dx.doi.org/10.1016/0377-0257(81)80020-1}.

\bibitem[Dek\'{e}e et~al.(1986)Dek\'{e}e, Carreau, and Mordarski]{Dekee1986}
D.~Dek\'{e}e, P.J. Carreau, and J.~Mordarski.
\newblock Bubble velocity and coalescence in viscoelastic liquids.
\newblock \emph{Chemical Engineering Science}, 41\penalty0 (9):\penalty0 2273
  -- 2283, 1986.
\newblock ISSN 0009-2509.
\newblock \doi{http://dx.doi.org/10.1016/0009-2509(86)85078-3}.

\bibitem[Gummalam et~al.(1988)Gummalam, Narayan, and Chhabra]{Gummalam1988}
S.~Gummalam, K.A. Narayan, and R.P. Chhabra.
\newblock Rise velocity of a swarm of spherical bubbles through a non-newtonian
  fluid: Effect of zero shear viscosity.
\newblock \emph{International Journal of Multiphase Flow}, 14\penalty0
  (3):\penalty0 361 -- 373, 1988.
\newblock ISSN 0301-9322.
\newblock \doi{http://dx.doi.org/10.1016/0301-9322(88)90050-X}.

\bibitem[Kee et~al.(1990)Kee, Chhabra, and Dajan]{Dekee1990}
D.~De Kee, R.P. Chhabra, and A.~Dajan.
\newblock Motion and coalescense of gas bubbles in non-newtonian polymer
  solutions.
\newblock \emph{Journal of Non-Newtonian Fluid Mechanics}, 37\penalty0
  (1):\penalty0 1 -- 18, 1990.
\newblock ISSN 0377-0257.
\newblock \doi{http://dx.doi.org/10.1016/0377-0257(90)80001-G}.

\bibitem[Manjunath and Chhabra(1992)]{Manjunath1992}
M.~Manjunath and R.P. Chhabra.
\newblock Free rise velocity of a swarm of spherical gas bubbles through a
  quiescent power law liquid.
\newblock \emph{International Journal of Engineering Science}, 30\penalty0
  (7):\penalty0 871 -- 878, 1992.
\newblock ISSN 0020-7225.
\newblock \doi{http://dx.doi.org/10.1016/0020-7225(92)90016-A}.

\bibitem[Miyahara and Yamanaka(1993)]{Miyahara1993}
Toshiro Miyahara and Shuichi Yamanaka.
\newblock Mechanics of motion and deformation of a single bubble rising through
  quiescent highly viscous newtonian and non-newtonian media.
\newblock \emph{Journal of Chemical Engineering of Japan}, 26\penalty0
  (3):\penalty0 297--302, 1993.
\newblock \doi{10.1252/jcej.26.297}.

\bibitem[Liu et~al.(1995)Liu, Liao, and Joseph]{Liu1995}
Y.~J. Liu, T.~Y. Liao, and D.~D. Joseph.
\newblock A two-dimensional cusp at the trailing edge of an air bubble rising
  in a viscoelastic liquid.
\newblock \emph{Journal of Fluid Mechanics}, 304:\penalty0 321--342, 12 1995.
\newblock ISSN 1469-7645.
\newblock \doi{10.1017/S0022112095004447}.
\newblock URL \url{http://journals.cambridge.org/article_S0022112095004447}.

\bibitem[Rodrigue et~al.(1996)Rodrigue, Kee, and Fong]{Rodrigue1996}
D.~Rodrigue, D.~De Kee, and C.F. Chan~Man Fong.
\newblock An experimental study of the effect of surfactants on the free rise
  velocity of gas bubbles.
\newblock \emph{Journal of Non-Newtonian Fluid Mechanics}, 66\penalty0
  (2):\penalty0 213 -- 232, 1996.
\newblock ISSN 0377-0257.
\newblock \doi{http://dx.doi.org/10.1016/S0377-0257(96)01486-3}.

\bibitem[Li et~al.(1997)Li, Mouline, Choplin, and Midoux]{Li1997}
H.Z. Li, Y.~Mouline, L.~Choplin, and N.~Midoux.
\newblock Chaotic bubble coalescence in non-newtonian fluids.
\newblock \emph{International Journal of Multiphase Flow}, 23\penalty0
  (4):\penalty0 713 -- 723, 1997.
\newblock ISSN 0301-9322.
\newblock \doi{http://dx.doi.org/10.1016/S0301-9322(97)00004-9}.

\bibitem[Chhabra(1998)]{Chhabra1998}
R.~P. Chhabra.
\newblock Rising velocity of a swarm of spherical bubbles in power law fluids
  at high reynolds numbers.
\newblock \emph{The Canadian Journal of Chemical Engineering}, 76\penalty0
  (1):\penalty0 137--140, 1998.
\newblock ISSN 1939-019X.
\newblock \doi{10.1002/cjce.5450760118}.
\newblock URL \url{http://dx.doi.org/10.1002/cjce.5450760118}.

\bibitem[Gauri and Koelling(1998)]{Gauri1999}
Vishal Gauri and W.~Kurt Koelling.
\newblock The motion of long bubbles through viscoelastic fluids in capillary
  tubes.
\newblock \emph{Rheologica Acta}, 38\penalty0 (5):\penalty0 458--470, 1998.
\newblock ISSN 1435-1528.
\newblock \doi{10.1007/s003970050197}.
\newblock URL \url{http://dx.doi.org/10.1007/s003970050197}.

\bibitem[Dubash and Frigaard(2007)]{Dubash2007Propagation}
N.~Dubash and I.A. Frigaard.
\newblock Propagation and stopping of air bubbles in carbopol solutions.
\newblock \emph{Journal of Non-Newtonian Fluid Mechanics}, 142\penalty0
  (1):\penalty0 123 -- 134, 2007.
\newblock ISSN 0377-0257.
\newblock \doi{http://dx.doi.org/10.1016/j.jnnfm.2006.06.006}.
\newblock URL
  \url{http://www.sciencedirect.com/science/article/pii/S0377025706001637}.

\bibitem[Tsamopoulos et~al.(2008)Tsamopoulos, Dimakopoulos, Chatzidai,
  Karapetsas, and Pavlidis]{Tsamopoulos2008Steady}
J.~Tsamopoulos, Y.~Dimakopoulos, N.~Chatzidai, G.~Karapetsas, and M.~Pavlidis.
\newblock Steady bubble rise and deformation in {Newtonian} and viscoplastic
  fluids and conditions for bubble entrapment.
\newblock \emph{Journal of Fluid Mechanics}, 601:\penalty0 123--164, 2008.
\newblock \doi{10.1017/S0022112008000517}.
\newblock URL
  \url{http://www.scopus.com/inward/record.url?eid=2-s2.0-42449160607&partnerID=40&md5=c188f7b8fcb9c9f30299e32bba9d8c77}.

\bibitem[Sikorski et~al.(2009)Sikorski, Tabuteau, and
  de~Bruyn]{Sikorski2009Motion}
Darek Sikorski, Herve Tabuteau, and John~R. de~Bruyn.
\newblock Motion and shape of bubbles rising through a yield-stress fluid.
\newblock \emph{Journal of Non-Newtonian Fluid Mechanics}, 159\penalty0
  (1):\penalty0 10 -- 16, 2009.
\newblock ISSN 0377-0257.
\newblock \doi{http://dx.doi.org/10.1016/j.jnnfm.2008.11.011}.
\newblock URL
  \url{http://www.sciencedirect.com/science/article/pii/S0377025708002140}.

\bibitem[Kishore et~al.(2013)Kishore, Nalajala, and Chhabra]{Kishore2013}
Nanda Kishore, V.~S. Nalajala, and Raj~P. Chhabra.
\newblock Effects of contamination and shear-thinning fluid viscosity on drag
  behavior of spherical bubbles.
\newblock \emph{Industrial \& Engineering Chemistry Research}, 52\penalty0
  (17):\penalty0 6049--6056, 2013.
\newblock \doi{10.1021/ie4003188}.

\bibitem[Tripathi et~al.(2015)Tripathi, Sahu, Karapetsas, and
  Matar]{Tripathi2015}
Manoj~Kumar Tripathi, Kirti~Chandra Sahu, George Karapetsas, and Omar~K. Matar.
\newblock Bubble rise dynamics in a viscoplastic material.
\newblock \emph{Journal of Non-Newtonian Fluid Mechanics}, \penalty0
  (0):\penalty0 article in press, 2015.
\newblock ISSN 0377-0257.

\bibitem[Kulkarni and Joshi(2005)]{Kulkarni2005}
Amol~A. Kulkarni and Jyeshtharaj~B. Joshi.
\newblock Bubble formation and bubble rise velocity in gas-liquid systems: A
  review.
\newblock \emph{Industrial \& Engineering Chemistry Research}, 44\penalty0
  (16):\penalty0 5873--5931, 2005.

\bibitem[Chhabra(2006)]{Chhabra2006}
R.P. Chhabra.
\newblock \emph{Bubbles, Drops, and Particles in Non-{Newtonian} Fluids}.
\newblock CRC Press, 2006.

\bibitem[Abbassi et~al.(2018)Abbassi, Besbes, Elhajem, Aissia, and
  Champagne]{Abbassi2018}
W.~Abbassi, S.~Besbes, M.~Elhajem, H.~Ben Aissia, and J.Y. Champagne.
\newblock Numerical simulation of free ascension and coaxial coalescence of air
  bubbles using the volume of fluid method (vof).
\newblock \emph{Computers \& Fluids}, 161:\penalty0 47--59, 2018.
\newblock ISSN 0045-7930.

\bibitem[Gumulya et~al.(2017)Gumulya, Utikar, Evans, Joshi, and
  Pareek]{Gumulya2017}
M.~Gumulya, R.P. Utikar, G.M. Evans, J.B. Joshi, and V.~Pareek.
\newblock Interaction of bubbles rising inline in quiescent liquid.
\newblock \emph{Chemical Engineering Science}, 166:\penalty0 1--10, 2017.
\newblock ISSN 0009-2509.

\bibitem[Watanabe and Sanada(2006)]{Watanabe2006}
Masao Watanabe and Toshiyuki Sanada.
\newblock In-line motion of a pair of bubbles in a viscous liquid.
\newblock \emph{Jsme International Journal Series B-fluids and Thermal
  Engineering}, 49:\penalty0 410--418, 2006.

\bibitem[Yuan and Prosperetti(1994)]{yuan_prosperetti_1994}
H.~Yuan and A.~Prosperetti.
\newblock On the in-line motion of two spherical bubbles in a viscous fluid.
\newblock \emph{Journal of Fluid Mechanics}, 278:\penalty0 325–349, 1994.

\bibitem[Legendre et~al.(2003)Legendre, Magnaudet, and
  Mougin]{legendre_magnaudet_mougin_2003}
Dominique Legendre, Jacques Magnaudet, and Guillaume Mougin.
\newblock Hydrodynamic interactions between two spherical bubbles rising side
  by side in a viscous liquid.
\newblock \emph{Journal of Fluid Mechanics}, 497:\penalty0 133–166, 2003.

\bibitem[Yu et~al.(2011)Yu, Yang, and Fan]{Yu2011LBM}
Zhao Yu, Hui Yang, and Liang-Shih Fan.
\newblock Numerical simulation of bubble interactions using an adaptive lattice
  boltzmann method.
\newblock \emph{Chemical Engineering Science}, 66\penalty0 (14):\penalty0
  3441--3451, 2011.
\newblock ISSN 0009-2509.
\newblock 10th International Conference on Gas–Liquid and
  Gas–Liquid–Solid Reactor Engineering.

\bibitem[Katz and Meneveau(1996)]{KATZ1996239}
J.~Katz and C.~Meneveau.
\newblock Wake-induced relative motion of bubbles rising in line.
\newblock \emph{International Journal of Multiphase Flow}, 22\penalty0
  (2):\penalty0 239--258, 1996.
\newblock ISSN 0301-9322.

\bibitem[Ruzicka(2000)]{RUZICKA20001141}
M.C Ruzicka.
\newblock On bubbles rising in line.
\newblock \emph{International Journal of Multiphase Flow}, 26\penalty0
  (7):\penalty0 1141--1181, 2000.
\newblock ISSN 0301-9322.

\bibitem[Hirt and Nichols(1981)]{Hirt1981}
C.~W Hirt and B.~D Nichols.
\newblock {Volume of Fluid (VOF) method for the dynamics of free boundaries}.
\newblock \emph{Journal of Computational Physics}, 39\penalty0 (1):\penalty0
  201--225, January 1981.
\newblock ISSN 00219991.

\bibitem[Rudman(1998)]{Rudman1998}
Murray Rudman.
\newblock {A volume-tracking method for incompressible multifluid flows with
  large density variations}.
\newblock \emph{International Journal for Numerical Methods in Fluids},
  28\penalty0 (2):\penalty0 357--378, August 1998.
\newblock ISSN 0271-2091.

\bibitem[Cummins et~al.(2005)Cummins, Francois, and Kothe]{Cummins2005}
Sharen~J. Cummins, Marianne~M. Francois, and Douglas~B. Kothe.
\newblock Estimating curvature from volume fractions.
\newblock \emph{Computers \& Structures}, 83\penalty0 (6-7):\penalty0 425 --
  434, 2005.
\newblock ISSN 0045-7949.

\bibitem[Francois et~al.(2006)Francois, Cummins, Dendy, Kothe, Sicilian, and
  Williams]{Francois2006}
Marianne~M. Francois, Sharen~J. Cummins, Edward~D. Dendy, Douglas~B. Kothe,
  James~M. Sicilian, and Matthew~W. Williams.
\newblock {A balanced-force algorithm for continuous and sharp interfacial
  surface tension models within a volume tracking framework}.
\newblock \emph{Journal of Computational Physics}, 213\penalty0 (1):\penalty0
  141--173, March 2006.
\newblock ISSN 00219991.

\bibitem[Wang and Tong(2010)]{Wang2010}
Zhaoyuan Wang and Albert~Y. Tong.
\newblock {A sharp surface tension modeling method for two-phase incompressible
  interfacial flows}.
\newblock \emph{International Journal for Numerical Methods in Fluids},
  64\penalty0 (7):\penalty0 709--732, September 2010.
\newblock ISSN 02712091.

\bibitem[Kumar et~al.(2015{\natexlab{b}})Kumar, Jin, and
  Vanka]{Kumar2015numerical}
Purushotam Kumar, Kai Jin, and S.~Pratap Vanka.
\newblock A multi-{GPU} based accurate algorithm for simulations of gas-liquid
  flows.
\newblock \emph{Computational Methods and Tools in Thermal Fluids Sciences},
  pages 1--17, 2015{\natexlab{b}}.

\bibitem[Jin et~al.(2016)Jin, Kumar, Vanka, and Thomas]{Jin2016mhd_bubble}
K.~Jin, P.~Kumar, S.~P. Vanka, and B.~G. Thomas.
\newblock {Rise of an argon bubble in liquid steel in the presence of a
  transverse magnetic field}.
\newblock \emph{Physics of Fluids}, 28\penalty0 (9), 2016.
\newblock ISSN 10897666.
\newblock \doi{10.1063/1.4961561}.

\bibitem[Rider and Kothe(1998)]{Rider1998}
William~J. Rider and Douglas~B. Kothe.
\newblock Reconstructing volume tracking.
\newblock \emph{Journal of Computational Physics}, 141\penalty0 (2):\penalty0
  112--152, April 1998.
\newblock ISSN 00219991.

\bibitem[Noh and Woodward(1976)]{Noh1976}
WF~Noh and P~Woodward.
\newblock {SLIC (simple line interface calculation)}.
\newblock In \emph{Proceedings of the Fifth International Conference}, pages
  330--340, Berlin, 1976. Springer-Verlag.

\bibitem[Li(1995)]{Li1995}
Jie Li.
\newblock {Calcul d'Interface Affine par Morceaux}(piecewise linear interface
  calculation).
\newblock \emph{C. R. Acad. Sci. Paris}, 320\penalty0 (8):\penalty0 391--396,
  1995.

\bibitem[Ashgriz and Poo(1991)]{Ashgriz1991}
N~Ashgriz and JY~Poo.
\newblock {FLAIR:} flux line-segment model for advection and interface
  reconstruction.
\newblock \emph{Journal of Computational Physics}, 93\penalty0 (2):\penalty0
  449--468, April 1991.
\newblock ISSN 00219991.

\bibitem[Renardy and Renardy(2002)]{Renardy2002}
Yuriko Renardy and Michael Renardy.
\newblock {PROST:} a parabolic reconstruction of surface tension for the
  volume-of-fluid method.
\newblock \emph{Journal of Computational Physics}, 183\penalty0 (2):\penalty0
  400--421, December 2002.
\newblock ISSN 00219991.

\bibitem[Sussman(2003)]{Sussman2003}
Mark Sussman.
\newblock {A second order coupled level set and volume-of-fluid method for
  computing growth and collapse of vapor bubbles}.
\newblock \emph{Journal of Computational Physics}, 187\penalty0 (1):\penalty0
  110--136, May 2003.
\newblock ISSN 00219991.

\bibitem[Sussman et~al.(2007)Sussman, Smith, Hussaini, Ohta, and
  Zhi-Wei]{Sussman2007}
M.~Sussman, K.M. Smith, M.Y. Hussaini, M.~Ohta, and R.~Zhi-Wei.
\newblock {A sharp interface method for incompressible two-phase flows}.
\newblock \emph{Journal of Computational Physics}, 221\penalty0 (2):\penalty0
  469--505, February 2007.
\newblock ISSN 00219991.

\bibitem[Gueyffier et~al.(1999)Gueyffier, Li, Nadim, Scardovelli, and
  Zaleski]{Gueyffier1999}
Denis Gueyffier, Jie Li, Ali Nadim, Ruben Scardovelli, and St\'{e}phane
  Zaleski.
\newblock {Volume-of-Fluid} interface tracking with smoothed surface stress
  methods for three-dimensional flows.
\newblock \emph{Journal of Computational Physics}, 152\penalty0 (2):\penalty0
  423--456, July 1999.
\newblock ISSN 00219991.

\bibitem[Vanka et~al.(2016)Vanka, Kumar, Jin, and Thomas]{Vanka2016Single}
{S. P.} Vanka, P.~Kumar, K.~Jin, and {B. G.} Thomas.
\newblock Single and multiphase flow computations on graphics processing units.
\newblock \emph{International Conference on Computational Methods for Thermal
  Problems}, \penalty0 (217349), 2016.
\newblock ISSN 2305-5995.

\bibitem[Kumar(2016)]{Kumar2016thesis}
Purushotam Kumar.
\newblock \emph{Development and application of Lattice Boltzmann and accurate
  volume of fluid numerical techniques on graphics processing units}.
\newblock PhD thesis, University of Illinois at Urbana-Champaign, 2016.

\bibitem[Horwitz et~al.(2012)Horwitz, Kumar, and Vanka]{horwitz2012simulations}
Jeremy Horwitz, Purushotam Kumar, and Pratap Vanka.
\newblock Simulations of multiphase flow in a t-junction and distributor
  header.
\newblock \emph{Bulletin of the American Physical Society}, 57, 2012.

\bibitem[Horwitz et~al.(2013)Horwitz, Kumar, and Vanka]{horwitz2013simulations}
Jeremy Horwitz, Purushotam Kumar, and Pratap Vanka.
\newblock Simulations of three-dimensional droplet deformation in a square-duct
  at moderate reynolds numbers.
\newblock \emph{Bulletin of the American Physical Society}, 58, 2013.

\bibitem[Kumar et~al.(2013)Kumar, Horwitz, and Vanka]{kumar2013three}
Purushotam Kumar, Jeremy Horwitz, and Surya Vanka.
\newblock A three-dimensional numerical study of immiscible droplet deformation
  in a right angle bend.
\newblock \emph{Bulletin of the American Physical Society}, 58, 2013.

\bibitem[Kumar et~al.(2019)Kumar, Jin, and Vanka]{Kumar2019AJKFluids}
Purushotam Kumar, Kai Jin, and Surya~Pratap Vanka.
\newblock {Numerical Simulation of a Gas Bubble Rising in Power-Law Fluids
  Using a Sharp Surface Force Implementation}.
\newblock \textit{Proceedings of the ASME-JSME-KSME 2019 8th Joint Fluids
  Engineering Conference. Volume 5: Multiphase Flow}. San Francisco,
  California, USA, July 28-August 01, 2019.

\bibitem[Horwitz et~al.(2014)Horwitz, Kumar, and Vanka]{Horwitz2014_lbm}
JAK Horwitz, P~Kumar, and SP~Vanka.
\newblock Three-dimensional deformation of a spherical droplet in a square duct
  flow at moderate reynolds numbers.
\newblock \emph{International journal of multiphase flow}, 67:\penalty0 10--24,
  2014.

\bibitem[Horwitz et~al.(2019)Horwitz, Vanka, and Kumar]{Horwitz2019AJKFluids}
Jeremy A.~K. Horwitz, S.~P. Vanka, and P.~Kumar.
\newblock Lbm simulations of dispersed multiphase flows in a channel: Role of a
  pressure poisson equation.
\newblock \textit{Proceedings of the ASME-JSME-KSME 2019 8th Joint Fluids
  Engineering Conference. Volume 5: Multiphase Flow}. San Francisco,
  California, USA, July 28-August 01, 2019.

\bibitem[{Vanka} et~al.(2015){Vanka}, {Jin}, {Kumar}, and
  {Thomas}]{Vanka2015APS_DFD}
Surya~Pratap {Vanka}, Kai {Jin}, Purushotam {Kumar}, and Brian {Thomas}.
\newblock {Rise of an argon bubble in liquid steel in the presence of a
  transverse magnetic field}.
\newblock \emph{Bulletin of the American Physical Society}, 60, 2015.

\bibitem[{Kumar} and {Vanka}(2022)]{Kumar2022APS_DFD}
Purushotam {Kumar} and Pratap {Vanka}.
\newblock {Dynamics of Argon Gas Bubble Pair Rising in Liquid Steel in the
  Presence of a Transverse Magnetic Field}.
\newblock \emph{Bulletin of the American Physical Society}, 67, 2022.

\bibitem[Kumar and Vanka(2023)]{kumar2023_MHD_inline_Bubble}
Purushotam Kumar and Surya~Pratap Vanka.
\newblock Dynamics of argon gas bubbles rising in liquid steel in the presence
  of transverse magnetic field.
\newblock \emph{arXiv preprint arXiv:2301.01797}, 2023.

\bibitem[{Vanka} et~al.(2016){Vanka}, {Kumar}, and {Jin}]{Vanka2016APS_DFD}
Pratap {Vanka}, Purushotam {Kumar}, and Kai {Jin}.
\newblock {Numerical Simulation of Turbulent Bubbly Flow in a Vertical Square
  Duct}.
\newblock \emph{Bulletin of the American Physical Society}, 61, 2016.

\bibitem[Kumar and Vanka(2021)]{kumar2021_bubbly_flow}
Purushotam Kumar and Surya~Pratap Vanka.
\newblock Large scale gpu based simulations of turbulent bubbly flow in a
  square duct.
\newblock \emph{arXiv preprint arXiv:2104.01636}, 2021.

\bibitem[Ohta et~al.(2003)Ohta, Iwasaki, Obata, and
  Yoshida]{Ohta2003_shear_thinning}
Mitsuhiro Ohta, Eiji Iwasaki, Eiji Obata, and Yutaka Yoshida.
\newblock A numerical study of the motion of a spherical drop rising in
  shear-thinning fluid systems.
\newblock \emph{Journal of Non-Newtonian Fluid Mechanics}, 116\penalty0
  (1):\penalty0 95 -- 111, 2003.
\newblock ISSN 0377-0257.
\newblock \doi{http://dx.doi.org/10.1016/j.jnnfm.2003.08.004}.

\bibitem[Zhang et~al.(2010)Zhang, Yang, and Mao]{Zhang2010}
Li~Zhang, Chao Yang, and Zai-Sha Mao.
\newblock Numerical simulation of a bubble rising in shear-thinning fluids.
\newblock \emph{Journal of Non-Newtonian Fluid Mechanics}, 165\penalty0
  (11-12):\penalty0 555 -- 567, 2010.
\newblock ISSN 0377-0257.
\newblock \doi{http://dx.doi.org/10.1016/j.jnnfm.2010.02.012}.

\bibitem[Ohta et~al.(2012)Ohta, Kimura, Furukawa, Yoshida, and
  Sussman]{Ohta2012Shear_thickening}
Mitsuhiro Ohta, Sachika Kimura, Tomohiro Furukawa, Yutaka Yoshida, and Mark
  Sussman.
\newblock Numerical simulations of a bubble rising through a shear-thickening
  fluid.
\newblock \emph{Journal of Chemical Engineering of Japan}, 45\penalty0
  (9):\penalty0 713--720, 2012.
\newblock \doi{10.1252/jcej.12we041}.

\bibitem[Carreau(1972)]{Carreau1972}
Pierre~J. Carreau.
\newblock Rheological equations from molecular network theories.
\newblock \emph{Transactions of The Society of Rheology}, 16\penalty0 (1),
  1972.

\end{thebibliography}

\end{document}